\documentclass[amsmath,amssymb,prb,twocolumn,showpacs]{revtex4}
\usepackage{color,graphicx}
\usepackage{epsfig}
\usepackage{amssymb}
\usepackage{tabularx}
\usepackage{bm}

\def\be{\begin{equation}}
\def\ee{\end{equation}}
\def\ba{\begin{eqnarray}}
\def\ea{\end{eqnarray}}

\begin{document}
\title{Impurity-induced bound states in iron-based superconductors with $s$-wave $\cos k_x\cdot\cos k_y$ pairing symmetry}
\author{Wei-Feng Tsai, Yan-Yang Zhang, Chen Fang, and Jiangping Hu}
\address{Department of Physics, Purdue University, West
Lafayette, Indiana 47907, USA}

\date{\today}
\newcommand{\br}{\mathbf{r}}
\newcommand{\brprime}{{\mathbf{r}^\prime}}
\newcommand{\bk}{\mathbf{k}}
\newcommand{\bkprime}{{\mathbf{k}^\prime}}

\begin{abstract}
Using both the self-consistent Bogoliubov-de Gennes formulation and non-self-consistent $T$-matrix approach, we perform a comprehensive investigation of  the in-gap bound states induced by a localized single impurity in iron-based superconductors.
We focus on studying signatures associated with the unconventional sign-changed $s$-wave pairing symmetry. For a non-magnetic impurity, we find that there are two in-gap bounds, symmetric with respect to zero energy, only in the sign-changed $s$-wave pairing state, not in the sign-unchanged $s$-wave state, due to  the existence of  non-trivial Andreev bound states caused by the sign change. For a magnetic impurity, we find that due to the breakdown of the local time-reversal symmetry, there exist only bound state solutions (with orbital degeneracy) carrying one of the electron-spin polarizations around the impurity. As increasing the scattering strength, the system undergoes a {\it quantum phase transition} (level crossing) from a spin-unpolarized ground state to a spin-polarized one. While the results for the magnetic impurity are qualitatively similar in both the sign-changed and sign-unchanged $s$-wave superconducting (SC) states, the bound states in the first case are more robust and there is no $\pi$ phase shift of the SC gap near the impurity in the strong scattering regime.
\end{abstract}

\maketitle

\section{Introduction}

Iron-based superconductors, a newly discovered family of superconductors with layered conducting planes,\cite{kamihara2008,takahashi2008,
Chen2008,Chenxh2008,wen2008} have stimulated enormously theoretical and experimental studies in the condensed matter community. This is not only because of the $T_c$ can be as high as 55K, but is also due to its striking similarity to the high-$T_c$ cuprates. One evident observation is that the undoped iron-based superconductors, as cuprates, exhibit antiferromagnetic order (though still metallic) and  the superconductivity occurs only when electrons or holes are sufficiently doped into the Fe layers. Due to the close proximity to the magnetism, the mechanism of the superconductivity is expected to be unconventional.\cite{Boeri2008,seo2008,Mazin2008a,Wang2008a,si,Cvetkovic2009,Mishra2009,Daghofer2008,Eremin2008}

As an important hint to unveil the mechanism of the superconductivity, the determination of an explicit superconducting (SC) gap structure is indeed essential. The basic electronic band structures of the iron-based superconductors have been predicted by the first-principle calculations\cite{Singh2008a,Ma2008a} that there are two hole Fermi pockets around $\Gamma$ point and two electron Fermi pockets around $M$ point in the first Brillouin zone (FBZ) with all five orbitals of an iron atom getting involved. As a result of this complexity, undoubtedly, nailing down right SC gap structure is still one of the most challenging issues in this rapidly growing field.

On the theoretical side, there have been many proposals for the possible pairing symmetries of iron pnictides, including nodeless or nodal SC order parameters.\cite{Mazin2009} Majority of studies so far, either from weak-coupling or strong-coupling approach, suggest an extended $s$-wave state (so called $s_\pm$ state), where the relative sign of SC order parameters changes between the hole and electron pockets (see Fig.~\ref{fig:FS}).\cite{seo2008,Mazin2008a,Yao08,Wang2008a} In particular, in a recent paper done by Seo {\it et al.},\cite{seo2008} it gives, in the proposed two-orbital exchange coupling model,\cite{twoorbital} an explicit $\cos k_x\cdot\cos k_y$ form of the pairing symmetry in momentum space, as long as two general conditions are satisfied: (i) the next nearest-neighbor (NNN) superexchange coupling $J_2$ dominates, and (ii) Fermi pockets are small near the aforementioned spots in the FBZ.
Furthermore, this simple form, $\cos k_x\cdot\cos k_y$, is in good agreement with the SC gaps measured by the angle-resolved photo-emission spectroscopy (ARPES).\cite{Ding2008a,Nakayama2008a,Hasan2008} Although this model does   overestimate  its insulating behavior near the undoped regime, this deficiency is irrelevant to determination for the properties of the SC state.  We will focus on this pairing state (for simplicity, $s_{\pm}$ state) within this model throughout this paper.

On the experimental side, however, there is still no census on the gap structure of the SC state. For instance, the ARPES results support a {\it fully-gapped} SC state,\cite{Ding2008a,Nakayama2008a,Hasan2008} consistent with point-contact Andreev spectroscopy\cite{Chen2008a} and  magnetic resonances measured by neutron scattering experiments,\cite{Christianson2008,Lumsden2008,Chi2008} while some penetration depth experiments,\cite{Hashimoto2008,Gordon2008,Prozorov2009} NMR  experiments\cite{Matano2008,Terasaki2009,Fukazawa2009,Mukuka2009} and other experiments\cite{Check2008} seemingly contradict with former interpretation. Even though it has been argued that $s_\pm$ state could partially reconcile these difficulties,\cite{Parish2008,Parker2008b,Laad2009,Seo2009a} it is still far from this SC order parameter being settled down. This, again, reflects an urgent call for a practical way to detect the SC gap structure, especially with sensitivity of measuring the sign change in the internal SC phase.

Despite of several theoretical works proposing various ways to detect the phase structure of the SC order parameter,\cite{Ghaemi2008,Tsai2008,Parker2008a,Wu2009,Zhangyy2009} achieving experimental realization reliably remains challenging. Since disorder is an intrinsic property in doped superconductors, a comprehensive study of impurity effects can also help indirectly probing SC order parameters. Here we propose tunneling measurements of impurity-induced states as a probe which is sensitive to the internal phase of the gap function between electron and hole Fermi pockets. Such sort of measurements has been proved invaluable in determining the nodal $d$-wave pairing symmetry of  the high-$T_c$ cuprates.\cite{Balatsky2006} In addition, the study of the impurity effects can also provide useful information about the SC gap structure and even uncover competing orders.\cite{ SEO2007,Seo2008a,CHEN2002, 
BENA2004,GHOSAL2005, KIVELSON2003, PODOLSKY2003}

In this paper, we perform a detailed investigation of the impurity-induced bound states in iron-based superconductors within a two-orbital exchange coupling model. By using both the self-consistent Bogoliubov-de Gennes (BdG) formulation and non-self-consistent $T$-matrix approach we find the following general features. (i) For the non-magnetic (intra-orbital) impurity potential, there exist two in-gap bounds, symmetric with respect to zero energy, only in the $s$-wave $\cos k_x\cdot\cos k_y$ pairing state, not in the sign-unchanged $s$-wave state. The basic physics of this result stems from  the emergence of  non-trivial Andreev bound state  within the SC gap due to the impurity scattering that destroys any {\it unconventional} (as opposed to the usual $s$-wave) phase assignment.
%Such sharp feature reflecting the fact of relative sign change in the gap structure should be detectable in the tunneling spectroscopy experiment.
(ii) For the magnetic (intra-orbital) impurity potential, due to the breakdown of the local time-reversal symmetry, there exist only bound state solutions (with orbital degeneracy) for one of the electron-spin polarizations around the impurity. As increasing the scattering strength, the system undergoes a {\it quantum phase transition} (level crossing) from spin-unpolarized ground state to spin-polarized one. Unlike the case with the non-magnetic impurity, the results in the $s$-wave $\cos k_x\cdot\cos k_y$ pairing state are qualitatively similar to those in the usual $s$-wave state. However, the sign-changed pairing state can sustain more robust bound state solutions without a $\pi$ phase shift of the SC gap near the impurity in the strong scattering regime.
%(iii) In addition, the presence of the extra non-magnetic/magnetic, inter-orbital potential, ...
The rich spectral features in our calculated energy- and space-dependent local density of states (LDOS) may be directly resolvable by future scanning tunnel microscope and assist to ultimately determine the  phase structure of the SC order parameter.

The organization of this paper is as follows. In Sec. II we briefly introduce the model and formalism we adopted.
%, including self-consistent BdG formulation and non-self-consistent T-matrix approach.
In Secs. III and IV we present our results for the cases of non-magnetic- and magnetic-impurity systems, respectively. Finally, some remarks are given in Sec. V and we conclude in Sec. VI.

\section{Model and formalism}
\begin{figure}[tbhp]
\begin{center}
\includegraphics[width=0.3\textwidth]{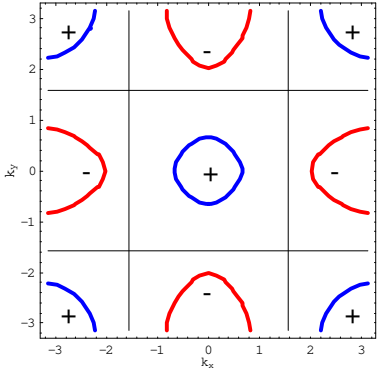}
\caption{Fermi surfaces of the two-orbital model representing iron pnictides at $\mu=1.65$ in the (unfolded) FBZ. The blue (red) curves correspond to the hole (electron) Fermi pockets. The $\pm$ sign within pockets indicate the relative sign change in the $\cos k_x\cdot\cos k_y$ SC order parameter. Also, the solid thin lines mark the nodal lines of it.}
\label{fig:FS}
\end{center}
\end{figure}

Our tight-binding microscopic Hamiltonian for the iron-based superconductors, describing iron atoms on a two-dimensional square lattice with two orbitals per site, is based on the so-called two-orbital exchange coupling model developed in Refs. \onlinecite{raghu2008,seo2008,Parish2008}. Explicitly, $H_{0}=H_t+H_{int}$,
where the non-interacting part reads
\ba
H_t &=& \sum_\mathbf{k}
\Psi^{\dagger}(\mathbf{k})\hat{h}_t(\bk)\Psi(\mathbf{k}), \nonumber \\
\hat{h}_t(\bk) &=&[(\epsilon_+(\mathbf{k}) -\mu)\sigma_0+\epsilon_-(\mathbf{k})\sigma_3+\epsilon_{xy}(\mathbf{k})\sigma_1]
\otimes\tau_3  \nonumber \\
\label{eq:h0}
\ea
with
$\Psi^{\dagger}(\mathbf{k})=(c^{\dagger}_{1,\mathbf{k},\uparrow},
c_{1,-\mathbf{k},\downarrow},c^{\dagger}_{2,\mathbf{k},\uparrow},
c_{2,-\mathbf{k},\downarrow})$ in Nambu spinor representation.
$c^\dagger_{\alpha,\bk,\sigma}$ creates an electron carrying momentum $\bk$ with orbital $\alpha$ ($\alpha=1,2$ for two degenerate ``$d_{xz}$'' and ``$d_{yz}$'' orbitals, respectively) and spin polarization $\sigma$. For a compact notation, we have made use of two sets of Pauli matrices, $\sigma_i$ and $\tau_i$, acting on orbital and particle-hole spaces, respectively, with $\sigma_0$ or $\tau_0$ the $2\times 2$ identity matrix. The matrix elements of $\hat{h}_t$, $\epsilon_+(\mathbf{k})=-(t_1+t_2)(\cos k_x+\cos k_y)-4t_3\cos k_x \cos k_y$,
$\epsilon_-(\mathbf{k})=-(t_1-t_2)(\cos k_x-\cos k_y)$, and $\epsilon_{xy}(\mathbf{k})=-4t_4\sin k_x\sin k_y$ are parametrized by four hopping amplitudes. The normal-state Fermi surfaces in the unfolded BZ can be reasonably produced by setting $t_1=-1.0,t_2=1.3$, and $t_3=t_4=-0.85$ (see Fig.~\ref{fig:FS}). For convenience, we have taken $|t_1|=1$ as our energy units, lattice constant $a\equiv 1$, and also have made $\mu=1.65$, which corresponds to electron density $n_e\approx 2.1$ per site. The interacting part contains several terms and can be expressed as
\ba
H_{int} &=& \sum_{\langle\br\brprime\rangle}\sum_{\alpha}
J_{1}(\mathbf{S}_{\alpha,\br}\cdot
\mathbf{S}_{\alpha,\brprime}-\frac{1}{4}n_{\alpha,\br}
n_{\alpha,\brprime}) \nonumber \\
&+& \sum_{\langle\langle\br\brprime\rangle\rangle}\sum_{\alpha}
J_{2}(\mathbf{S}_{\alpha,\br}\cdot
\mathbf{S}_{\alpha,\brprime}-\frac{1}{4}n_{\alpha,\br}
n_{\alpha,\brprime}) \nonumber \\
&+& \cdots,
\ea
where $\mathbf{S}_{\alpha,\br}=\frac{1}{2}c^\dagger_{\alpha,\br,\sigma}
\vec{\sigma}_{\sigma\sigma^\prime}c_{\alpha,\br,\sigma^\prime}$ and $n_{\alpha,\br}$ are the local spin and density operators with orbital $\alpha$. $\langle \br\brprime\rangle$ and $\langle\langle \br\brprime\rangle\rangle$ denote nearest-neighbor (NN) and NNN pairs of sites, respectively. ``$\cdots$'' represent our ignored orbital crossing exchange coupling and Hund's coupling terms, which are argued by one of us in Ref. \onlinecite{seo2008} to be unimportant on determining the pairing symmetry of the SC state in this model.

In this paper, we will focus on the $s$-wave $\cos k_x\cdot\cos k_y$ pairing symmetry in iron-based superconductors and neglect uncompetitive/subleading pairing symmetries such as $s$-wave $\cos k_x+\cos k_y$ and $d$-wave $\cos k_x-\cos k_y$ (as shown in Ref.~\onlinecite{seo2008}) by setting exchange coupling $J_1=0$. Also, we will assume that the low-energy physics of the system may reliably be captured by the mean-field approximation as long as the pairing interaction is small compared to the bandwidth. By defining the local $s_{\pm}$-wave pairing amplitude for orbital $\alpha$,
\be
\Delta_{\alpha}(\br,\br+\delta)=-J_2\langle c_{\alpha,\br,\downarrow}c_{\alpha,\br+\delta,\uparrow}
\rangle
\ee
with $\delta=\pm\hat{x}\pm\hat{y}$ indicating NNN pairing, the mean-field Hamiltonian of $H_0$ is then written as
\be
H^{MF}_{0}=H_t+\frac{1}{4}\sum_{\br,\delta,\alpha}
[\Delta^*_{\alpha}(\br,\br+\delta)c_{\alpha,\br,\downarrow}
c_{\alpha,\br+\delta,\uparrow}+h.c.]. \label{eq:mfH}
\ee

Furthermore, as a useful comparison, we shall consider the ``sign-unchanged'' onsite $s$-wave symmetry as well in order to extract the non-triviality of the ``sign-changed'' $s$-wave symmetry. In this case, we replace the interacting term by
$H_{int}=-|U|\sum_{\alpha,\br}n_{\alpha,\br,\uparrow}
n_{\alpha,\br,\downarrow}$ and make use of the following mean-field decoupling instead,
\be
-|U|\sum_{\alpha,\br,\sigma}\bar{n}_{\alpha,\br,\bar{\sigma}}n_{\alpha,\br,\sigma}
+\sum_{\alpha,\br}[\Delta^*_{s\alpha}(\br)c_{\alpha,\br,\downarrow}
c_{\alpha,\br,\uparrow}+h.c.],
\ee
where
\be
\Delta_{s\alpha}(\br)=-|U|\langle c_{\alpha,\br,\downarrow}c_{\alpha,\br,\uparrow}
\rangle,\quad \bar{n}_{\alpha,\br,\sigma}=\langle n_{\alpha,\br,\sigma}\rangle.
\ee

The interaction between the conduction electrons in the superconductor and a single non-magnetic impurity on site $\br_I$ can be written as
\be
H_{nimp}=V_{I}\sum_{\alpha,\sigma}c^\dagger_{\alpha,\br_I,\sigma}c_{\alpha,\br_I,\sigma}
+V_{I}^\prime\sum_{\alpha\neq\alpha^\prime,\sigma}c^\dagger_{\alpha,\br_I,\sigma}
c_{\alpha^\prime,\br_I,\sigma}, \label{eq:nimp}
\ee
where $V_{I}$ ($V_{I}^\prime$) represents the intra-orbital (inter-orbital) scattering strength. On the other hand, the scattering from a static (classical) magnetic impurity with magnetic moment $\vec{s}$ is given by
\be
H_{mimp}=J_{I}\sum_{\alpha}\mathbf{S}_{\alpha,\br_I}\cdot\vec{s}
+\frac{J^\prime_{I}}{2}\sum_{\alpha\neq\alpha^\prime,\sigma}
c^\dagger_{\alpha,\br_I,\sigma}
\vec{\sigma}_{\sigma\sigma^\prime}c_{\alpha^\prime,\br_I,\sigma^\prime}
\cdot\vec{s}, \label{eq:mimp}
\ee
where $J_{I}$ ($J_{I}^\prime$) represents the intra-orbital (inter-orbital) magnetic scattering strength. Note that due to spin-rotational symmetry, one can choose the coordinate system for the spin degrees of freedom such that $z$ axis points in the direction of $\vec{s}$. Consequently, it is sufficient to keep only $z$-component in Eq.~(\ref{eq:mimp}) hereafter.

\subsection{Self-consistent Bogoliubov-de Gennes formulation}
In order to treat the impurity scattering problem and a spatial variation in the superconducting order parameter on equal footing, we will mainly use self-consistent BdG formulation,\cite{Xiang95,Ghosal01} which has been proved beneficial in gaining real space information, to demonstrate our results of investigation.

Within this formulation, we diagonalize the {\it quadratic}, mean-field Hamiltonian (\ref{eq:mfH}) plus impurity term [either Eq.~(\ref{eq:nimp}) or (\ref{eq:mimp})] through the BdG equation,
\be
\left(\begin{array}{cccc}
\hat{K}_{1\sigma} & \hat{K}_{12\sigma} & \hat{\Delta}_1 & 0 \\
\hat{K}_{12\sigma} & \hat{K}_{2\sigma} & 0 & \hat{\Delta}_2  \\
\hat{\Delta}^*_1 & 0 & -\hat{K}^*_{1\bar{\sigma}} & -\hat{K}^*_{12\bar{\sigma}} \\
0 & \hat{\Delta}^*_2 & -\hat{K}^*_{12\bar{\sigma}} & -\hat{K}^*_{2\bar{\sigma}}
\end{array}
\right)\left(\begin{array}{c}
u_{1\br\sigma}^n \\
u_{2\br\sigma}^n \\
v_{1\br\bar{\sigma}}^n \\
v_{2\br\bar{\sigma}}^n \\
\end{array}
\right)=E_n\left(\begin{array}{c}
u_{1\br\sigma}^n \\
u_{2\br\sigma}^n \\
v_{1\br\bar{\sigma}}^n \\
v_{2\br\bar{\sigma}}^n \\
\end{array}
\right) \label{eq:bdg}
\ee
with $n$th eigenvalue $E_n$, and the operators in the matrix above obey
\ba
\hat{K}_{1\sigma}u^n_{1\br\sigma} &=& -t_1u^n_{1\br\pm\hat{x}\sigma}-t_2u^n_{1\br\pm\hat{y}\sigma}
-t_3\sum_{\delta}u^n_{1\br+\delta\sigma} \nonumber \\
&+&[(V_{I}+\sigma J_Is_z/2)\delta_{\br,\br_I}-\mu]u^n_{1\br\sigma}, \nonumber \\
\hat{K}_{2\sigma}u^n_{2\br\sigma} &=& -t_2u^n_{2\br\pm\hat{x}\sigma}-t_1u^n_{2\br\pm\hat{y}\sigma}
-t_3\sum_{\delta}u^n_{2\br+\delta\sigma} \nonumber \\
&+&[(V_{I}+\sigma J_Is_z/2)\delta_{\br,\br_I}-\mu]u^n_{2\br\sigma}, \nonumber
\ea
\ba
\hat{K}_{12\sigma}u^n_{1\br\sigma}&=& -t_4\sum_{\delta}e^{i \mathbf{Q}\cdot \delta}u^n_{2\br+\delta\sigma}+(V^\prime_{I}+\sigma J^\prime_Is_z/2)\delta_{\br,\br_I}u^n_{2\br\sigma}   \nonumber \\
\hat{\Delta}_{\alpha}v^n_{\alpha\br\sigma} &=& \frac{1}{4}\sum_{\delta}
\Delta_{\alpha}(\br,\br+\delta)v^n_{\alpha\br+\delta\sigma},\,\text{(similar to $u^n_{\alpha\br\sigma}$)}  \nonumber
\ea
where $\sigma=\pm$ correspond to spin up/down, $\delta$ are NNN vectors, and $\mathbf{Q}=(\frac{\pi}{2},\frac{\pi}{2})$. The relation between quasi-particle operators $\gamma$ and electron operators is
$c_{\alpha,\br,\sigma}=\sum_{n}(u^n_{\alpha\br\sigma}\gamma_{\alpha,n,\sigma}
-\sigma v^{n*}_{\alpha\br\sigma}\gamma^\dagger_{\alpha,n,\bar{\sigma}})$, and hence combining with the definition of $s$-wave $\cos k_x\cdot\cos k_y$ SC order parameter, this gives rise to the following self-consistent conditions,
\ba
\Delta_{\alpha}(\br,\br+\delta)&=&\frac{J_2}{2}\sum_{n}
(u^n_{\alpha\br\uparrow}v^{n*}_{\alpha\br+\delta\downarrow}
+u^n_{\alpha\br\downarrow}v^{n*}_{\alpha\br+\delta\uparrow}) \nonumber \\
&\times & \tanh \frac{E_n}{2k_B T}. \label{eq:selfcon_s2}
\ea
For onsite $s$-wave pairing, we instead have
\ba
\Delta_{s\alpha}(\br)&=&\frac{|U|}{2}\sum_{n}(u^n_{\alpha\br\uparrow}v^{n*}_{\alpha\br\downarrow}
+u^n_{\alpha\br\downarrow}v^{n*}_{\alpha\br\uparrow})\tanh \frac{E_n}{2k_B T},
\nonumber \\
\bar{n}_{\alpha,\br,\sigma} &=& \sum_{n}
|v^n_{\alpha\br\sigma}|^2[1-f(E_n)]+\sum_n|u^n_{\alpha\br\sigma}|^2f(E_n),
\nonumber\\  
\label{eq:selfcon_s}
\ea
where $f(E)$ is the Fermi distribution function. Notice that the summation here is only over those eigenstates with positive eigenvalues due to the symmetry property of the BdG equation in the whole {\it spin space}: If $(u^n_{1\uparrow},u^n_{2\uparrow},v^n_{1\downarrow},v^n_{2\downarrow},
u^n_{1\downarrow},u^n_{2\downarrow},v^n_{1\uparrow},
v^n_{2\uparrow})^t$ is an eigenfunction of the equation with eigenvalue $E_n$, then $(v^{n*}_{1\uparrow},v^{n*}_{2\uparrow},-u^{n*}_{1\downarrow},-u^{n*}_{2\downarrow},
-v^{n*}_{1\downarrow},-v^{n*}_{2\downarrow},u^{n*}_{1\uparrow},
u^{n*}_{2\uparrow})^t$ is also an eigenfunction with eigenvalue $-E_n$.

Unless otherwise stated, we always perform our computations on a finite lattice of $N=32\times 32$ sites with periodic boundary conditions at temperature $k_BT=0.03$. We obtain the resulting quasi-particle spectrum by repeatedly diagonalizing BdG Eq.~(\ref{eq:bdg}) and iteration of the pairing amplitudes according to self-consistency condition (\ref{eq:selfcon_s2}) [or Eq.~(\ref{eq:selfcon_s}) for onsite $s$-wave] until sufficient accuracy is achieved ({\it e.g.} the relative error of the pairing amplitudes is less than 1\%). We choose suitable $J_2=8$ and $U=2.56$ such that the ratio of the SC coherence peak, $\Delta_{coh}\approx 0.4$, to the maximum bandwidth of the bands, $W_{max}=12$, is around 0.033. With this choice, we can restrict the SC coherence length $\xi\sim\hbar v_F/\Delta_{coh}\lesssim 4a$, as suggested in experiments. Finally, it is practically useful to note that for the case of non-magnetic impurity, the subindex $\sigma$ becomes unimportant for $u$ and $v$, and we can save the computation effort by cutting the spin space half, {\it i.e.}, keeping only index $\sigma=\uparrow$ in Eq. (\ref{eq:bdg}).

\begin{figure}[tbh]
\begin{center}
\includegraphics[width=0.4\textwidth]{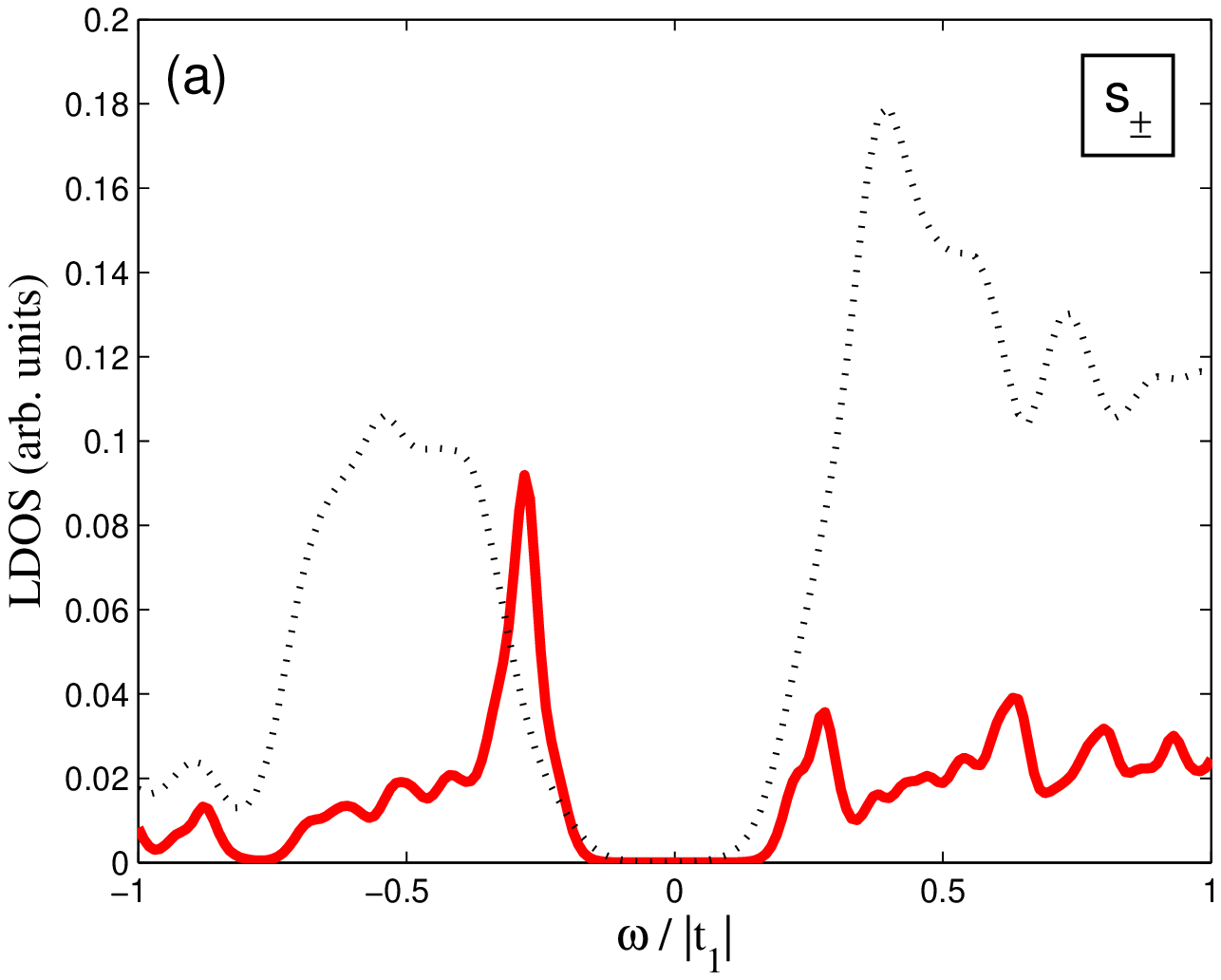}
\includegraphics[width=0.4\textwidth]{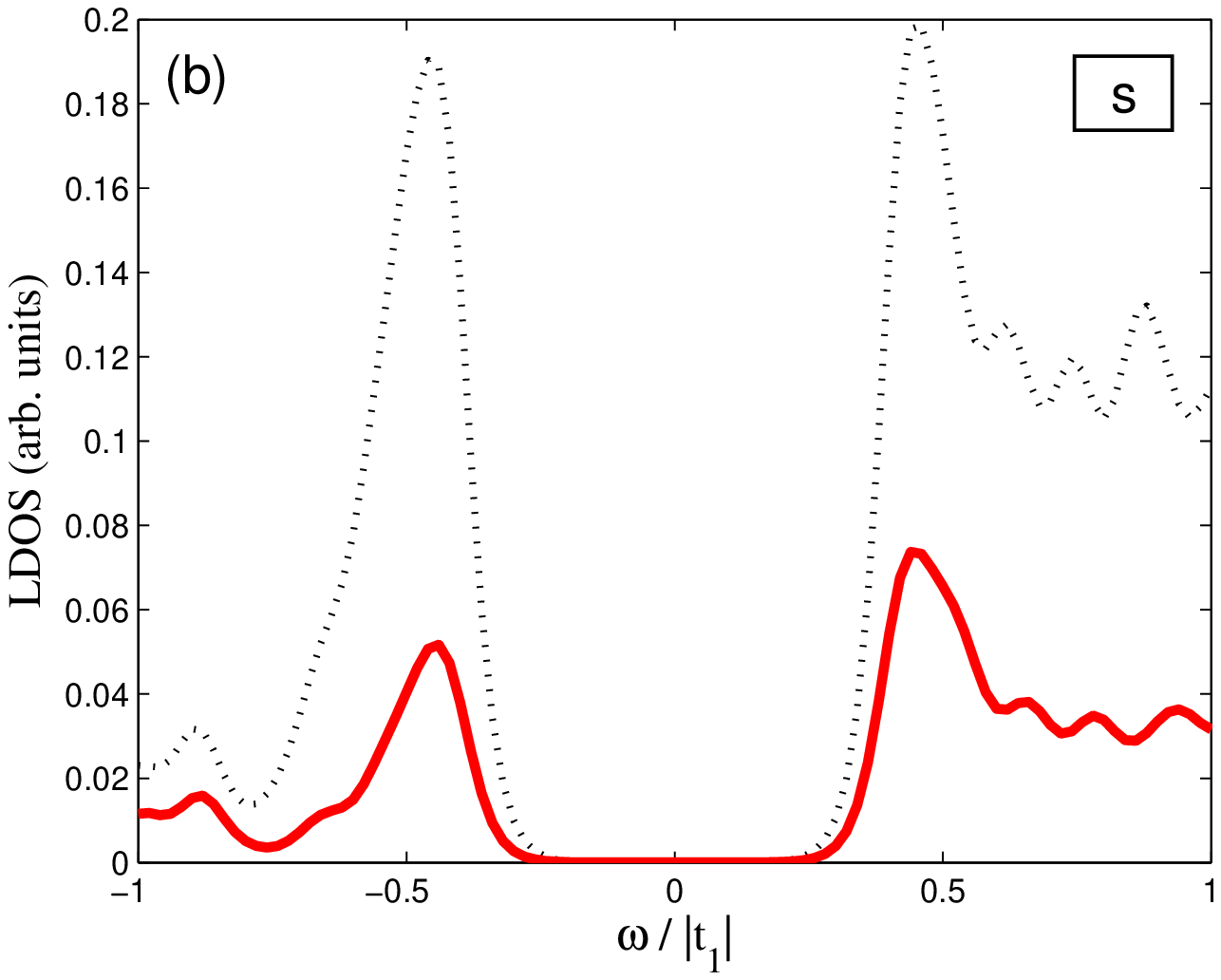}
\caption{(Color online) LDOS (red curve) at the nonmagnetic impurity site, $\br_I=(16,16)$, for (a) the $s_{\pm}$-wave pairing state and (b) the onsite $s$-wave pairing state with $V_I=4$. The dotted black curves represent the LDOS at the same position (renormalized by a factor of 4) in a clean system for comparison. ($\Delta_{coh}\approx 0.4$)}
\label{fig:nimpLDOS}
\end{center}
\end{figure}

\subsection{Non-self-consistent $T$-matrix approximation}
By assuming that the relaxation of the superconducting order parameter is negligible, we greatly simplify our impurity-scattering problem and make the physics more transparent. As we will see later, the validity of using $T$-matrix approach\cite{Balatsky2006,Wang04} to capture qualitatively right physics is justified by comparing with the results from the BdG formulation. In addition, the main difference between a ``self-consistent'' treatment and a ``non-self-consistent'' treatment in this approach is the inclusion of (normal/anomalous) self-energy corrections to the bare electron Green's function or not. Basically, these corrections are proportional to the impurity concentration and hence can be reasonably ignored for a single-impurity problem, as in our case here. However, the price we have to pay for this simplification is the loss of complete information around the impurity site.

Let us start with a mean-field Hamiltonian in the SC state, $H_0^{MF}=\sum_\mathbf{k}
\Psi^{\dagger}(\mathbf{k})\hat{h}(\bk)\Psi(\mathbf{k})$, where $\hat{h}(\bk)=\hat{h}_{t}(\bk)
+\Delta(\mathbf{k})\sigma_0\otimes\tau_1$, with notations defined before in the momentum space. Note that we have taken a suitable gauge choice to make SC order parameter real and set for $s_{\pm}$-wave pairing,  $\Delta_{\alpha}(\mathbf{k})=\Delta(\bk)=\Delta_0\cos k_x\cos k_y$, while for onsite $s$-wave pairing, $\Delta_{s\alpha}(\mathbf{k})=\Delta(\bk)=\Delta_{s0}$. In the same compact notation, now the impurity potential can be unified as $H_{imp}=\sum_{\mathbf{k},\mathbf{k}^\prime}\Psi^\dagger(\mathbf{k})\hat{V}
\Psi(\mathbf{k}^\prime)$ with $\hat{V}=V_{\mu\nu}\sigma_{\mu}\otimes\tau_{\nu}$ (no summation over $\mu$ and $\nu$), where different types of impurity scattering problems are related by $V_{03}=V_I,V_{13}=V_I^\prime,V_{00}=J_Is_z/2$, and $V_{10}=J_I^\prime s_z/2$ (otherwise, $V_{\mu\nu}=0$).

Defining $\tilde{\omega}=\omega+i0^+$, the Green's function for a clean SC system is given by
\be
G^{0}(\bk,\tilde{\omega})=[\tilde{\omega}I_4-\hat{h}(\bk)]^{-1}\equiv
\left(\begin{array}{cc}
  G^{0}_{11}(\bk,\tilde{\omega}) & G^{0}_{12}(\bk,\tilde{\omega}) \\
G^{0}_{21}(\bk,\tilde{\omega}) & G^{0}_{22}(\bk,\tilde{\omega})
\end{array}
\right), \label{eq:G0}
\ee
where $G^0_{\alpha\beta}$ is a $2\times 2$ matrix acting on the particle-hole space. The full Green's function in the single-impurity problem within $T$-matrix approximation is then written as
\be
G(\bk,\bkprime,\tilde{\omega})=G^{0}(\bk,\tilde{\omega})\delta_{\bk,\bkprime}+
G^{0}(\bk,\tilde{\omega})T(\bk,\bkprime,\tilde{\omega})
G^{0}(\bkprime,\tilde{\omega}) \nonumber
\ee
with the whole impurity-induced effect contained only in the $T$-matrix.
Standard perturbation theory gives
\begin{eqnarray}
T(\tilde{\omega})&=&\hat{V}+\hat{V} g^0(\tilde{\omega})\hat{V}\nonumber+\hat{V} g^0(\tilde{\omega})\hat{V} g^0(\tilde{\omega})\hat{V}+\ldots\\
&=&[I_4-\hat{V}g^0(\tilde{\omega})]^{-1}\hat{V}, \label{eq:tmatrix}
\end{eqnarray}
where $g^0(\tilde{\omega})=\int\frac{d^2k}{(2\pi)^2}G^0(\mathbf{k},\tilde{\omega})$.
These equations allow us to determine the solutions of the impurity-induced bound states via Det($T^{-1}$)=0 in the sub-gap regime, $|\omega|<|\Delta(\bk)|$.

\section{Nonmagnetic impurity}
We begin with our discussion on the effect of nonmagnetic (scalar) impurity in $s_{\pm}$-wave superconductors. It is well known that nonmagnetic impurities in a single-band, isotropic $s$-wave superconductor are not hard pair breakers and hence the impurity-induced spectral feature lies essentially at the gap edge. \cite{Fetter65,Shiba73,Balatsky2006} However, as we will see later, it is not the case for a $s_{\pm}$-wave superconductor due to its non-trivial SC phase structure in the momentum space. Any potential scattering between hole and electron Fermi pockets may destroy the phase assignment and leads to the formation of nontrivial in-gap bound states.

\begin{figure}[tbh]
\begin{center}
\includegraphics[width=0.4\textwidth]{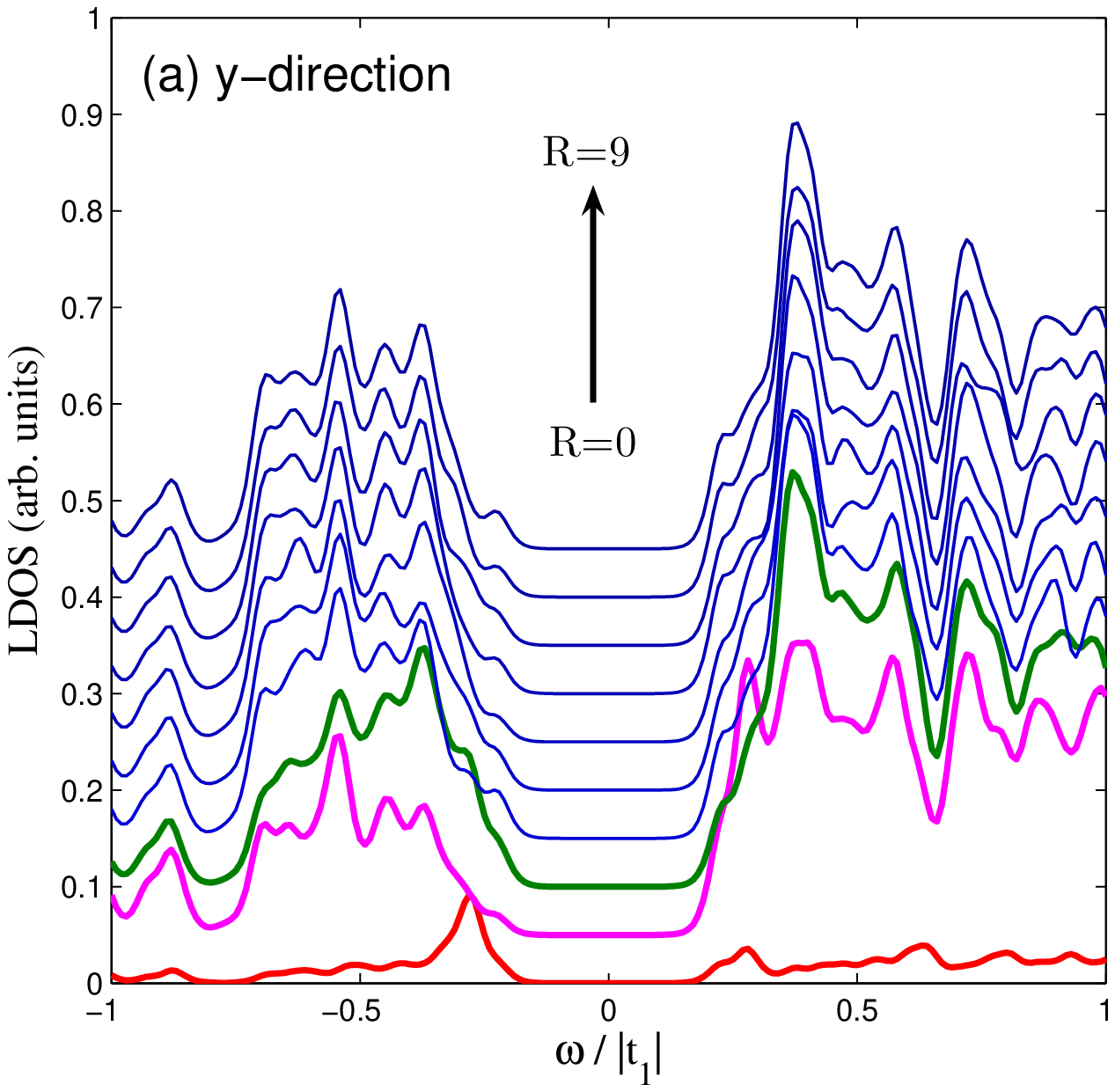}
\includegraphics[width=0.4\textwidth]{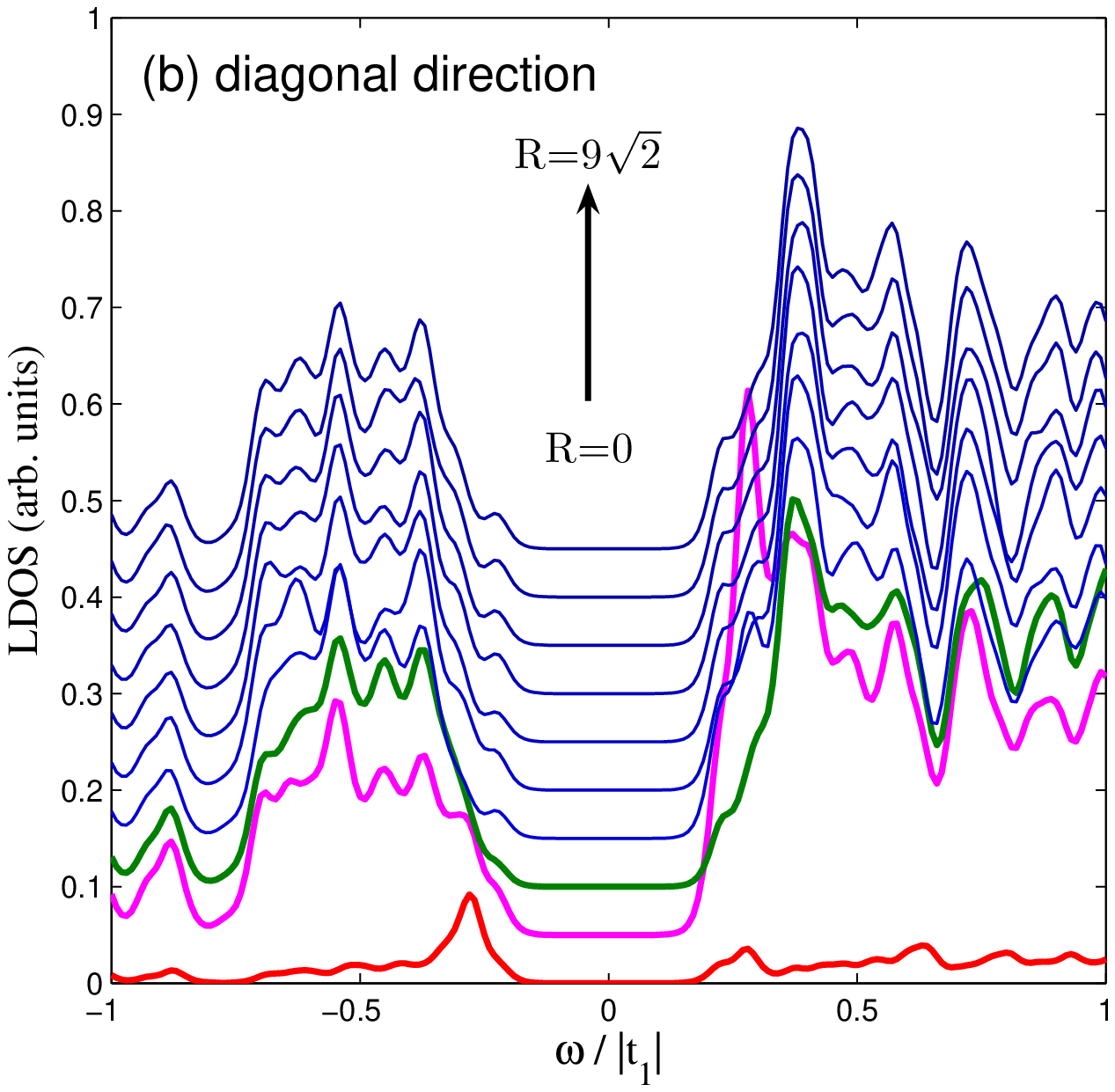}
\caption{(Color online) LDOS as a function of energy $\omega$ and the distance $R$ away from the nonmagnetic impurity along (a) +$y$ direction and (b) diagonal direction. The red, pink, and green curves represent LDOS at the impurity site, its first neighbor, and its second neighbor along $y$ or diagonal direction, respectively. All curves with different $R$s are shifted by 0.05 along vertical axis with each other. Note that the parameters are the same as those used in Fig.~\ref{fig:nimpLDOS} and the self-consistent pairing potential at $\br_I$ is around 0.17.}
\label{fig:bs_Rmap}
\end{center}
\end{figure}

Consider first the case of intra-orbital potential scattering, where $V_I\neq 0$ and $V_I^\prime=J_I=J_I^\prime=0$. For a $s_{\pm}$-wave superconductor, we show in Fig.~\ref{fig:nimpLDOS}(a) the LDOS, which is obtained via
\be
N_{\sigma}(\omega,\br)=-\frac{1}{N}\sum_{n,\alpha}[|u^n_{\alpha\br\sigma}|^2
f^\prime(E_n-\omega)+|v^n_{\alpha\br\bar{\sigma}}|^2f^\prime(E_n+\omega)], \label{eq:ldos}
\ee
then summing over spin $\sigma$ at the impurity site, and compare it with the one without any impurity (dotted curve). For the demonstration purpose, we choose a moderate scattering strength, $V_I=4$. The spectroscopic signature of bound-state solutions is clearly seen as two peaks, symmetric with respect to zero energy and within the SC coherence peak $\Delta_{coh}\approx 0.4$, in the LDOS at the impurity site. Furthermore, the weaker spectral weight at positive energy and the stronger one at negative energy arise from the absence of particle-hole symmetry in the system. The presence of such in-gap bound states in $s_{\pm}$ pairing state is indeed in sharp contrast to the similar problem in the sign-unchanged $s$-wave pairing state, where no peak can be found within the SC gap as shown in Fig.~\ref{fig:nimpLDOS}(b). The localized nature of the impurity-induced states within the gap is further proved by showing the LDOS as a function of energy $\omega$ and the distance $R=|\br-\br_I|$ off the impurity position along certain directions in Fig.~\ref{fig:bs_Rmap}. As $R$ is away from the impurity site by one or two lattice constants, the peaks disappear quickly and the LDOS recovers back to the shape of the bulk DOS.

\begin{figure}[bth]
\begin{center}
\includegraphics[width=0.238\textwidth]{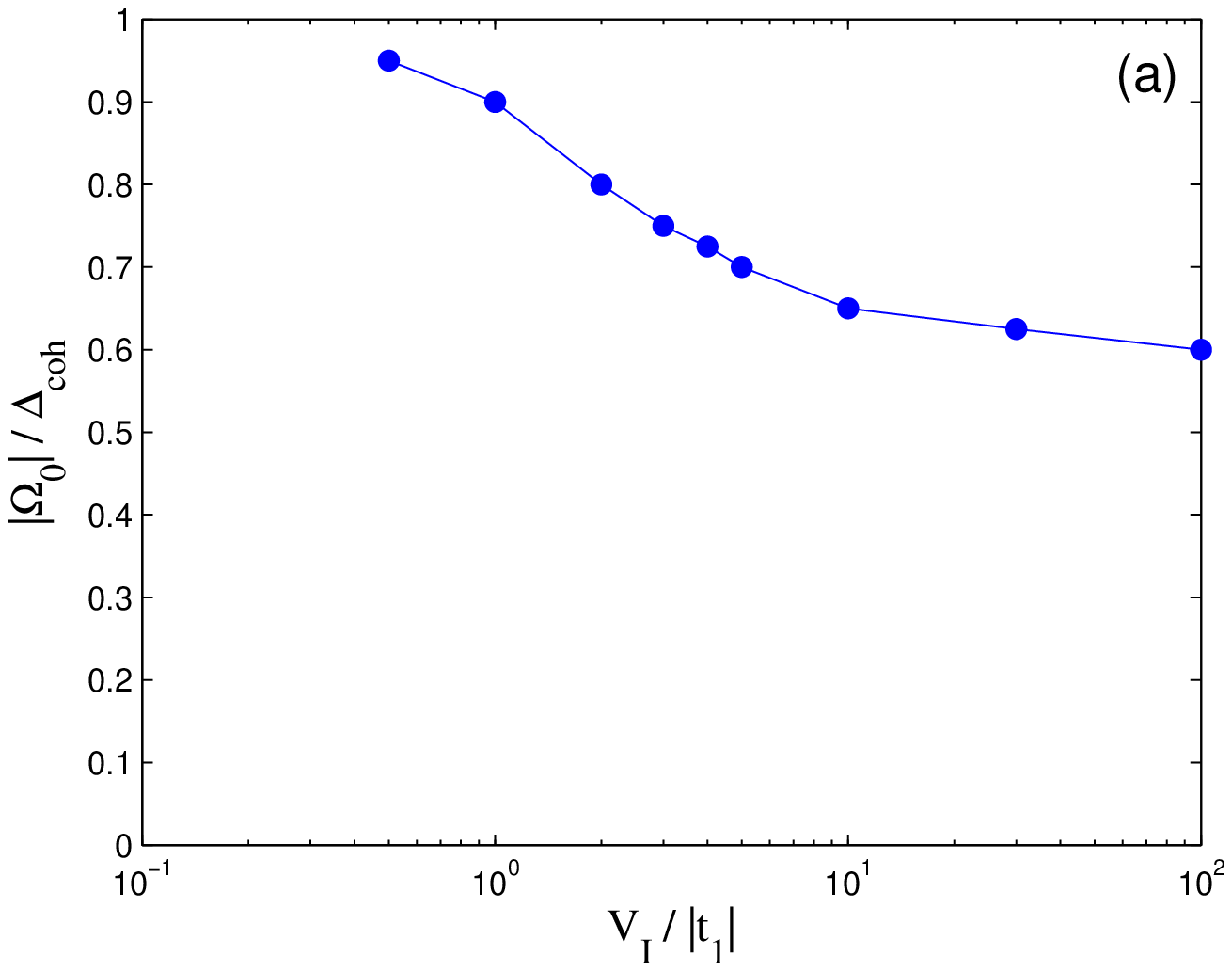}
\includegraphics[width=0.238\textwidth]{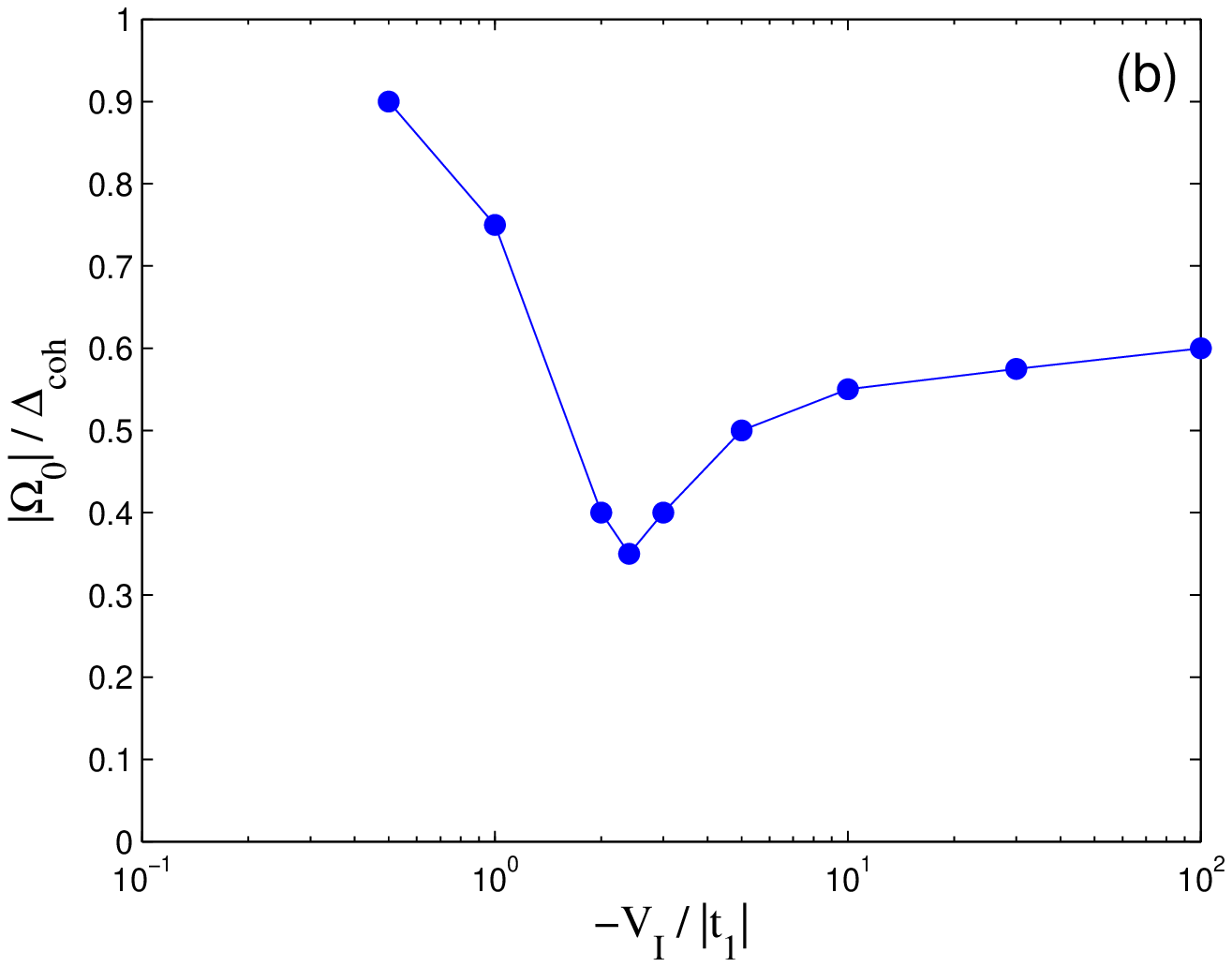}
\caption{Bound-state energy as a function of non-magnetic, intra-orbital impurity scattering strength: (a) $V_I>0$ and (b) $V_I<0$. Note that the $x$ axis is drawn in logarithmic scale. ($\Delta_{coh}= 0.4$)}
\label{fig:BSenergy}
\end{center}
\end{figure}

Above results can be qualitatively understood by (non-self-consistent) $T$-matrix approach. Combining $g^0(\tilde{\omega})$ for the $s_{\pm}$-wave pairing state derived in Appendix and the inverse of the scattering matrix $\hat{V}^{-1}$, the inverse of the $T$-matrix in Eq.~(\ref{eq:tmatrix}) is given by
\be
T^{-1}(\tilde{\omega})=V_{I}^{-1}\sigma_0\otimes[\tau_3+c_n(\frac{\tilde{\omega}}
{\sqrt{\Delta_{coh}^2-\tilde{\omega}^2}}\tau_0-\gamma_0\tau_3)], \label{eq:tinverse}
\ee
where we have defined the dimensionless scattering strength $c_n= V_{I}\pi\rho_0$ [$\pi\rho_0\sim\mathcal{O}(1)$ in the iron pnictides, and $\gamma_0< 0$ is related to the used energy cutoff here (see Appendix)]. It is important to realize that as $\omega^2<\Delta_{coh}^2$, Im$(T^{-1})\rightarrow 0$. Thus, true bound states at {\it real} $\omega$ could be found by the condition Det$(T^{-1})=0$, leading to the bound-state energy,
\be
\Omega_0=\pm\Delta_{coh}\frac{|1-c_n\gamma_0|}{\sqrt{(1-c_n\gamma_0)^2+c_n^2}}.
\label{eq:nimpbsenergy}
\ee
This is in contrast to the case of the nodal $d$-wave pairing, where we usually get {\it virtual} bound states at {\it complex} $\omega$.\cite{Balatsky2006} Note that for each solution in Eq.~(\ref{eq:nimpbsenergy}), it is doubly degenerate due to orbital degeneracy. Also, as sharply opposed to the $s_{\pm}$-wave pairing, there are no in-gap bound states found for the onsite $s$-wave state, quite consistent with the results shown in Figs.~\ref{fig:nimpLDOS}(a) and \ref{fig:nimpLDOS}(b). In fact, this distinct feature comes from the (nearly) absence of $\tau_1$ component in $T^{-1}(\tilde{\omega})$, which is proportional to $\Delta_{coh}$ in the sign-unchanged $s$-wave pairing.

\begin{figure} [htb]
\begin{center}
\includegraphics[scale=0.45]{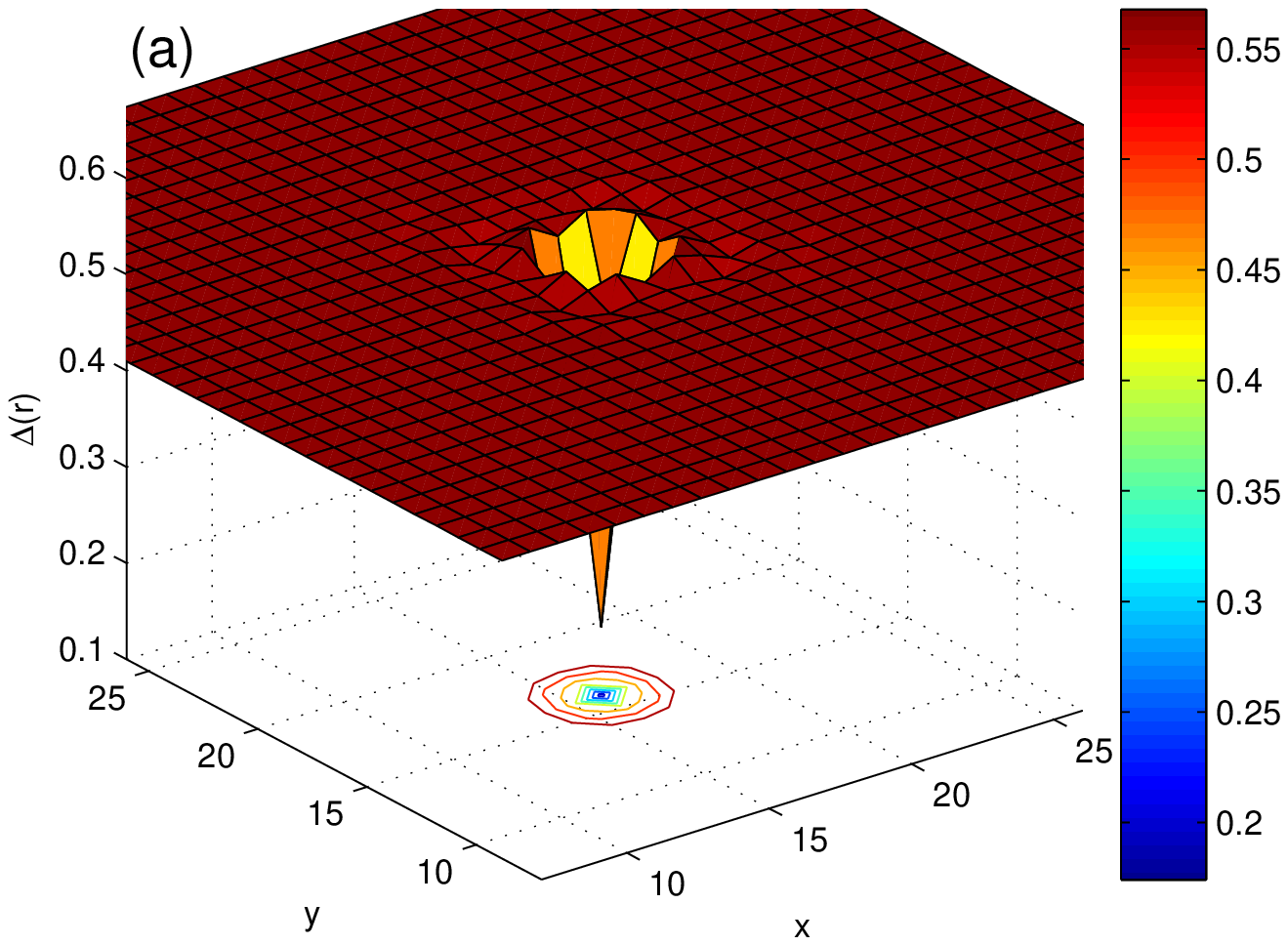}
\includegraphics[scale=0.45]{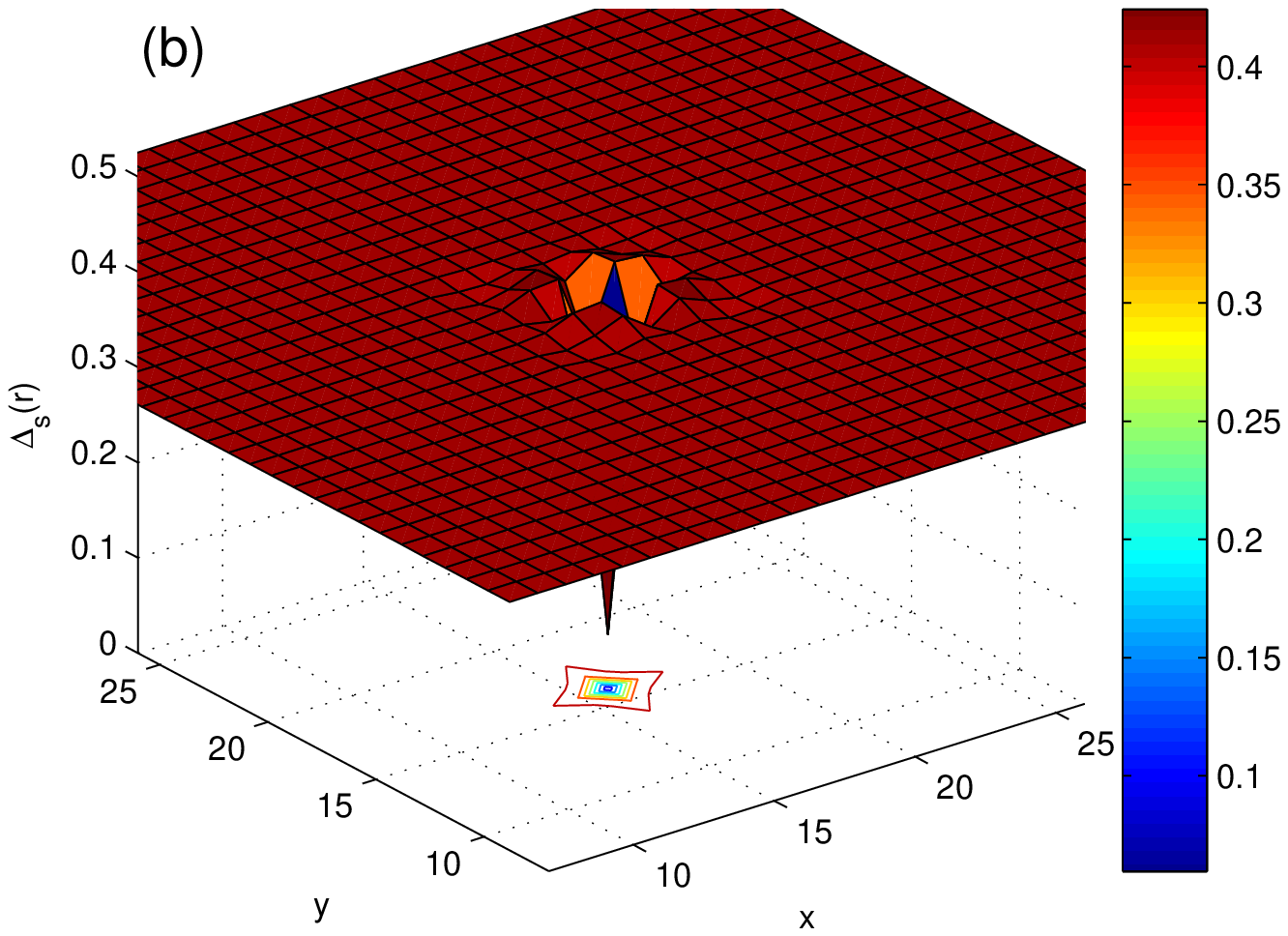}
\end{center}
\caption{(Color online) The spatial dependence of the SC gaps for (a) the $s_{\pm}$-wave pairing symmetry and (b) the onsite $s$-wave pairing symmetry in the presence of a non-magnetic impurity at $\br_I$=(16,16).}
\label{fig:SCgap}
\end{figure}

In addition, by increasing the scattering strength $V_I>0$, the bound-state solutions for the $s_{\pm}$-wave pairing state, as estimated by $T$-matrix approach, evolves from the gap edge in the weak scattering limit to $\pm\Delta_{coh}\gamma_0/\sqrt{\gamma_0^2+1}$ in the unitary scattering limit.
However, the qualitative behavior changes greatly when increasing the magnitude of the {\it negative} scattering strength. From Eq.~(\ref{eq:nimpbsenergy}), reversing the sign of $c_n$ tells us that there must be a minimum bound state energy, occurring at a critical $V_I<0$. (It does not reach zero energy in our system because a small $\tau_1$ component in $T^{-1}$ would appear practically due to imperfect cancellation of the $\Delta$ terms on electron and hole Fermi pockets and the SC gap relaxation should be taken into account as well.)
The numerical results, as seen in Figs.~\ref{fig:BSenergy}(a) and \ref{fig:BSenergy}(b), indeed follow what we have discussed from $T$-matrix consideration.

\begin{figure} [htbp]
\begin{center}
\includegraphics[width=0.4\textwidth]{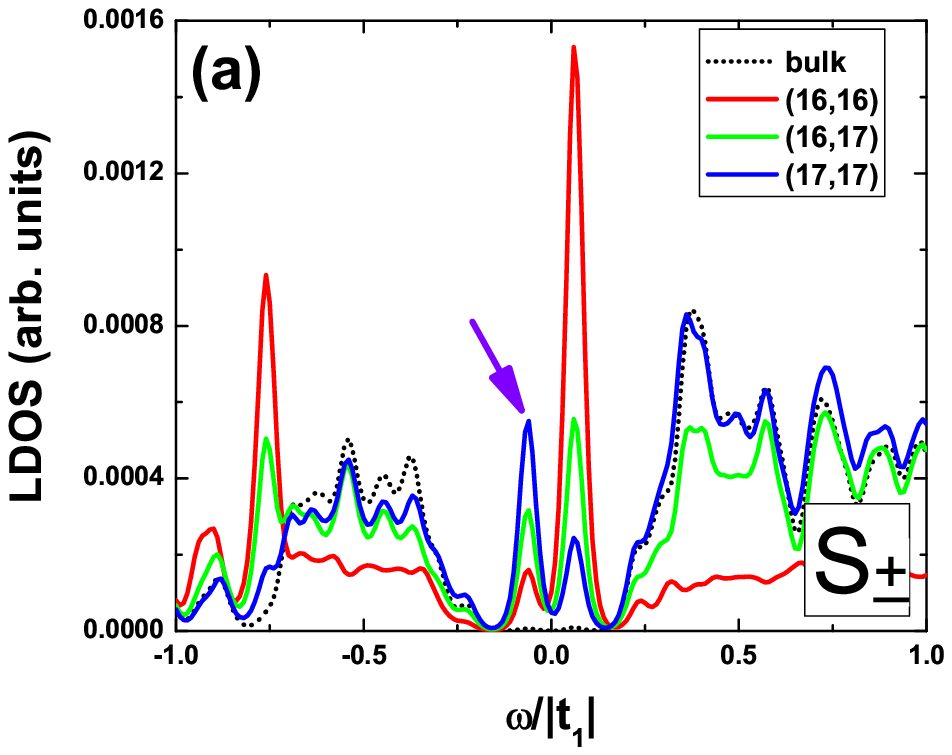}
\includegraphics[width=0.4\textwidth]{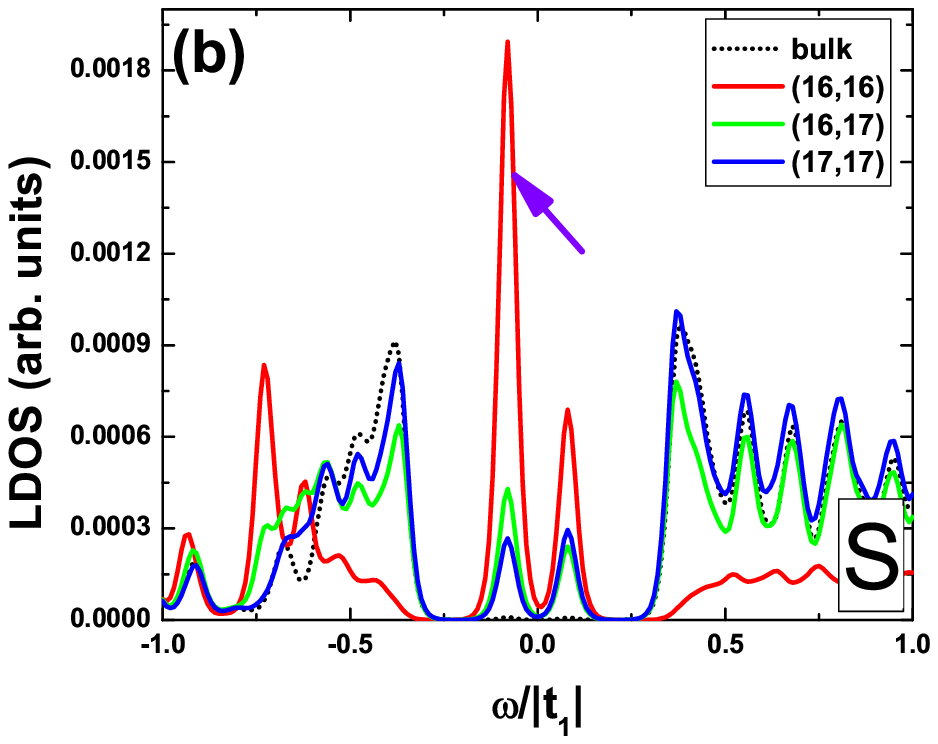}
\end{center}
\caption{(Color online) LDOS as a function of energy $\omega$ (red curves) in the presence of a magnetic impurity at $\br_I=(16,16)$ for (a) $s_{\pm}$-wave with $J_Is_z/2=2$ and (b) onsite $s$-wave with $J_Is_z/2=2.2$.
The black, dotted curves represent the bulk DOS normalized by the number of sites $N$ for useful comparison ($\Delta_{coh}\approx 0.4$).}
\label{fig:GraphLDOS}
\end{figure}

Next, we consider the change in the SC gap function caused by a single nonmagnetic-impurity scattering. Figure~\ref{fig:SCgap}(a) shows the self-consistent SC pairing potential $\Delta(\br)=\sum_{\alpha,\delta}\Delta_{\alpha}(\br,\br+\delta)/8$ with $s_{\pm}$-wave symmetry.
Since the impurity potential is short-ranged, the SC gap changes largely in the vicinity of the impurity site and recovers soon to its maximum value away from the impurity. It is clear  that there are two relevant length scales controlling the behavior. The shorter one associates with the range of the impurity potential, in which the SC gap is strongly suppressed due to much smaller electron population. The other scale is comparable to the SC coherence length $\xi\lesssim 4a$, in which the SC gap is weakly oscillating. This oscillation simply indicates the competition between the impurity potential and the SC pairing potential. Another subtle feature in Fig.~\ref{fig:SCgap}(a) is that the contour line of $\Delta$ is anisotropic and in roughly diamond shape round the impurity site. This should be due to the fact that the DOS from  $d_{xz}$ and $d_{yz}$ orbitals is most likely dominated by the (elliptic) electron pockets around $(0,\pi)$ and $(\pi,0)$ in this model, causing the anisotropy of the SC coherence length. This feature may only slightly depend on the specific form of the pairing symmetry since similar anisotropy of the spatial dependence of the onsite $s$-wave SC gap function, $\Delta_s(\br)=\sum_{\alpha}\Delta_{s\alpha}(\br)/2$, is also seen in Fig.~\ref{fig:SCgap}(b). Interestingly, we also observe that the spatial distribution of LDOS at $\Omega_0$, $N(\Omega_0,\br)$, respect the same anisotropy (not shown).

As a final remark in this section, we comment on the case when there exists small component of the inter-orbital, nonmagnetic-impurity scattering, {\it i.e.}, $\hat{V}=(V_I\sigma_0+V_I^\prime\sigma_1)\otimes\tau_3$. We sketch the analysis briefly below by using the $T$-matrix approximation. The easiest way to consider this problem is to make a unitary transformation in the orbital space such that the transformed scattering matrix becomes $(V_I\sigma_0+V_I^\prime\sigma_3)\otimes\tau_3$. Also, the transformed $g^0(\tilde{\omega})$ is the same as before due to the fact that it is diagonal in the orbital space. Thus, we now simply deal with new impurity potentials $V_I\pm V_I^\prime$ separately within each (transformed) orbital. The direct consequence is simply the breakdown of the orbital degeneracy such that each bound-state energy $\Omega_0$ splits into two. This is still a distinguishable feature from the $s$-wave pairing state, in which essentially no sharp peaks in LDOS at the impurity site when $|\omega|<\Delta_{coh}$.

\section{(Classical) Magnetic impurity}
We now turn to the discussion on the effect of magnetic impurities in the $s_{\pm}$ pairing state. The magnetic impurities, the pair breakers of the Cooper pairs, are known to induce in-gap bound states in the conventional nodeless superconductors.\cite{Yu65,Shiba68,Rusinov69} These bound-state solutions are usually localized near the impurity, possibly with nontrivial spin configuration around it, and may dramatically modify the ground-state properties of the system once the magnetic interaction is much stronger than the condensation energy, $\Delta_{coh}$.

\begin{figure}[htbp]
\begin{center}
\includegraphics[width=0.42\textwidth]{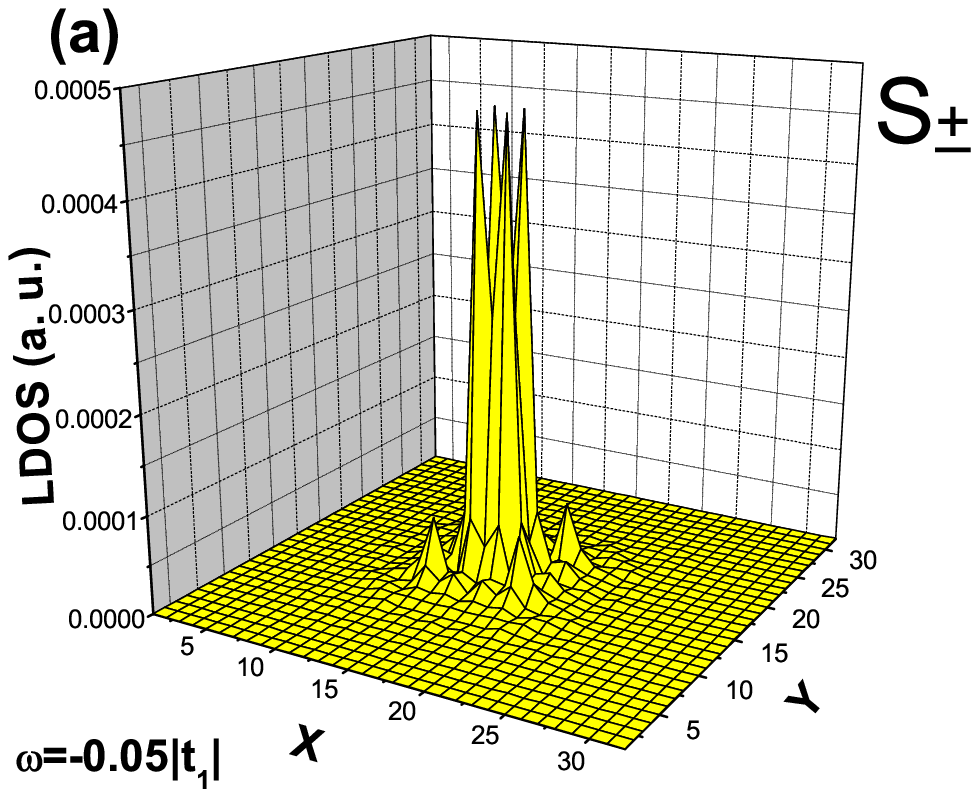}
\includegraphics[width=0.39\textwidth]{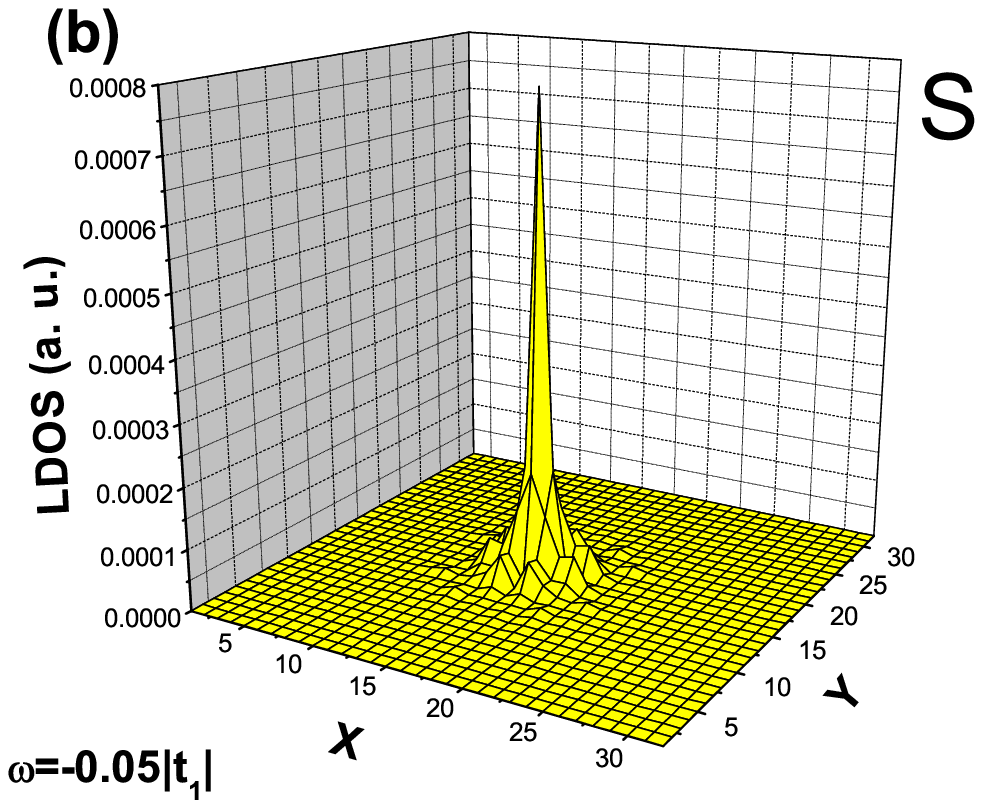}
\end{center}
\caption{(Color online) The spatial distribution of LDOS for
(a) $s_{\pm}$-wave at $\Omega_0=-0.05$, and (b) onsite $s$-wave at $\Omega_0=-0.05$. They correspond to the arrows indicated in Figs.~\ref{fig:GraphLDOS}(a) and \ref{fig:GraphLDOS}(b), respectively.} \label{fig:BS_LDOSR}
\end{figure}

To investigate possible features in our target pairing state, we consider the case of intra-orbital, purely magnetic impurity scattering, where $J_I\neq 0$ while $V_I=V_I^\prime=J_I^\prime=0$. Examining the LDOS spectrum at the impurity site in Fig.~\ref{fig:GraphLDOS}(a), one can immediately see that there are two sharp peaks, symmetric with respect to zero energy within the SC gap ($\Delta_{coh}\approx 0.4$), indicating the presence of the bound-state solutions. The asymmetric spectral weights for the peaks are again due to the breakdown of the particle-hole symmetry in the system. However, the presence of in-gap peaks in the LDOS is also observed in the onsite $s$-wave pairing state as shown in Fig.~\ref{fig:GraphLDOS}(b), except for the reversed magnitudes of the spectral weights on the two peaks. This subtle difference may not be considered as a general feature since the asymmetry of the spectral weights depends on the position and the scattering strength. Furthermore, as we examine the LDOS spectra away from the impurity, the peak positions do not change and their spectral weights  decay rapidly after a few lattice constants comparable to the SC coherence length $\xi$. In Fig.~\ref{fig:BS_LDOSR}(a), the spatial distribution of LDOS at fixed bound-state energy, $N(\pm\Omega_0,\br)$ [see Eq.~(\ref{eq:ldos})], further confirms our observation. While in Fig.~\ref{fig:BS_LDOSR}(b), we find no essential difference for the sign-unchanged $s$-wave pairing state, except that the spectral weights  are more concentrated on the impurity site. For energies outside the SC gap, the spatial distribution becomes much more extended with only a suppressed region around the impurity (not shown here), as one can expect.

\begin{figure} [htbp]
\begin{center}
\includegraphics[width=0.238\textwidth]{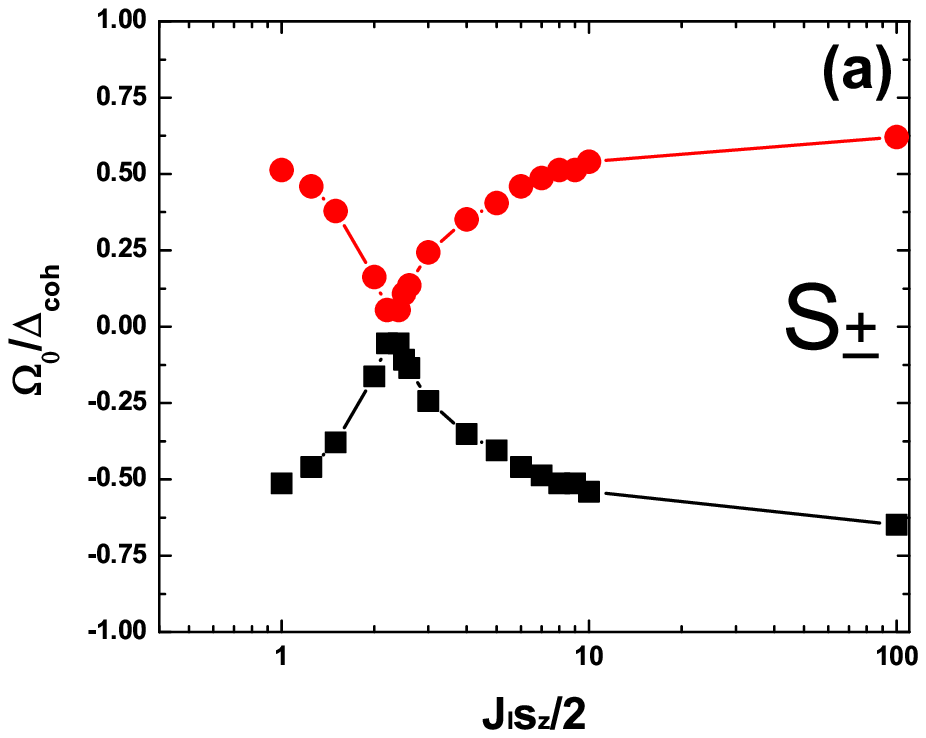}
\includegraphics[width=0.238\textwidth]{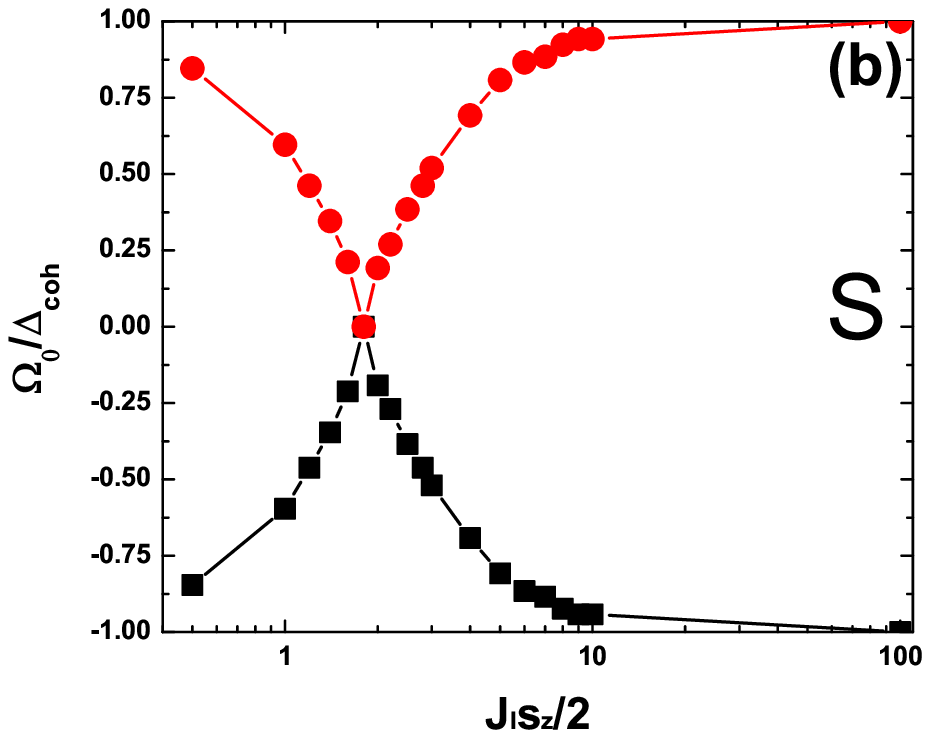}
\end{center}
\caption{(Color online) The energy of bound states $\Omega_0$ as a
function of the effective magnetic impurity moment $J_Is_z/2$ for (a)
$s_{\pm}$-wave pairing and (b) onsite $s$-wave pairing. ($\Delta_{coh}\approx 0.4$)}
\label{fig:LocationBS}
\end{figure}

Similar to the nonmagnetic-impurity problem discussed in the last section, the $T$-matrix approach may assist us to understand the properties of the magnetic-induced bound states more clearly. For the $s_{\pm}$-wave pairing symmetry, the inverse of the $T$-matrix is obtained via the replacement of $V_I^{-1}$ and $c_n$ in Eq.~(\ref{eq:tinverse}) by the inverse of the effective magnetic moment, $(s^{eff})^{-1}$, and the dimensionless magnetic scattering strength, $c_m=s^{eff}\pi\rho_0$ ($s^{eff}\equiv J_Is_z/2$). Det$(T^{-1})=0$ is satisfied at,\cite{comment_tau1}
\be
\Omega_0=\Delta_{coh}\frac{1\mp |c_m\gamma_0|}{\sqrt{(1\mp|c_m\gamma_0|)^2+c_m^2}}.
\label{eq:mimpBS}
\ee
For $c_m>0,\gamma_0< 0$, we will ignore the second bound-state solution above [``+'' sign in Eq.~(\ref{eq:mimpBS})] since it is very close the gap edge as $c_m\ll 1$, while its magnitude approaches to that of the first solution as $c_m\gg 1$. Therefore, it is hardly discernible in our numerical LDOS results. The first bound-state solution in Eq.~(\ref{eq:mimpBS}) is to be compared with the one obtained from the case with onsite $s$-wave symmetry, where the in-gap bound state occurs at
\be
\Omega_0=\Delta_{coh}\frac{1-c_m^2(1+\gamma_0^2)}
{\sqrt{[1-c_m^2(1+\gamma_0^2)]^2+4c_m^2}}. \label{eq:mimpBS_s}
\ee
Now, we see that in both pairing symmetries, the presence of the in-gap bound states is consistent with our numerical results shown in Figs.~\ref{fig:GraphLDOS}(a) and \ref{fig:GraphLDOS}(b). In addition, one should notice that, first, each solution is doubly degenerate due to orbital degeneracy; second, although the quasi-particle bound-state energy $\Omega_0$ appears {\it not} symmetric with respect to zero energy, the resonance peaks in LDOS are indeed symmetric since each quasi-particle state has its particle and hole components at $-|\Omega_0|$ and $|\Omega_0|$ separately.\cite{Balatsky2006} So far, in some sense, the magnetic-impurity effect could not be a good probe to distinguish the sign-changed $s$-wave symmetry from the sign-unchanged one. Nevertheless, there are still ``subtle'' features possessed only by the $s_{\pm}$-wave pairing state, as we will explain next.

\begin{figure} [htbp]
\begin{center}
\includegraphics[width=0.4\textwidth]{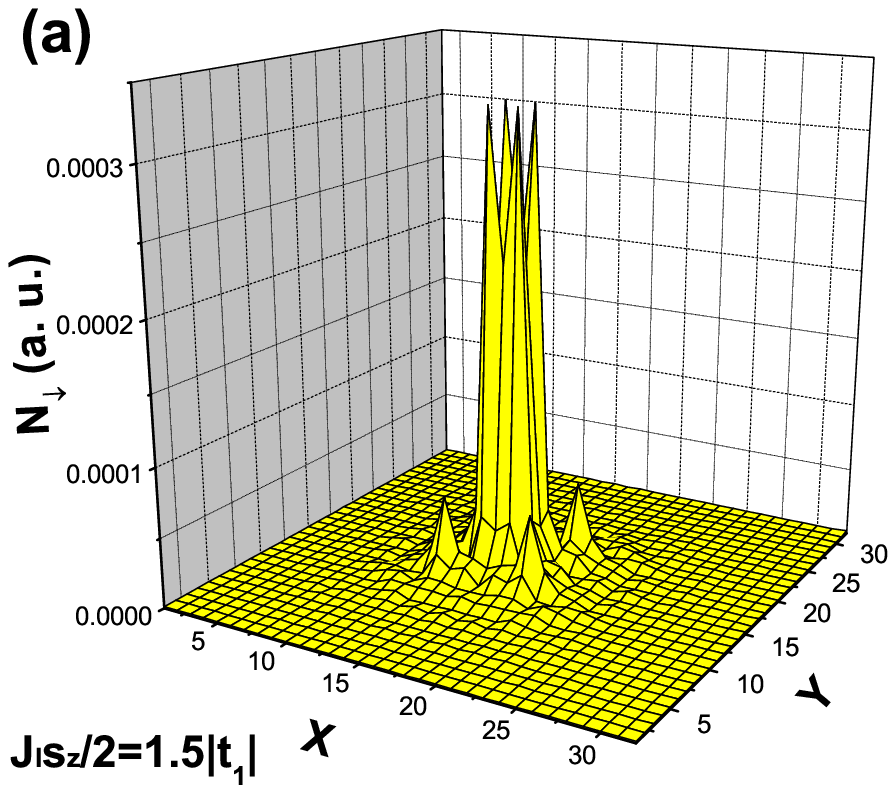}
\includegraphics[width=0.4\textwidth]{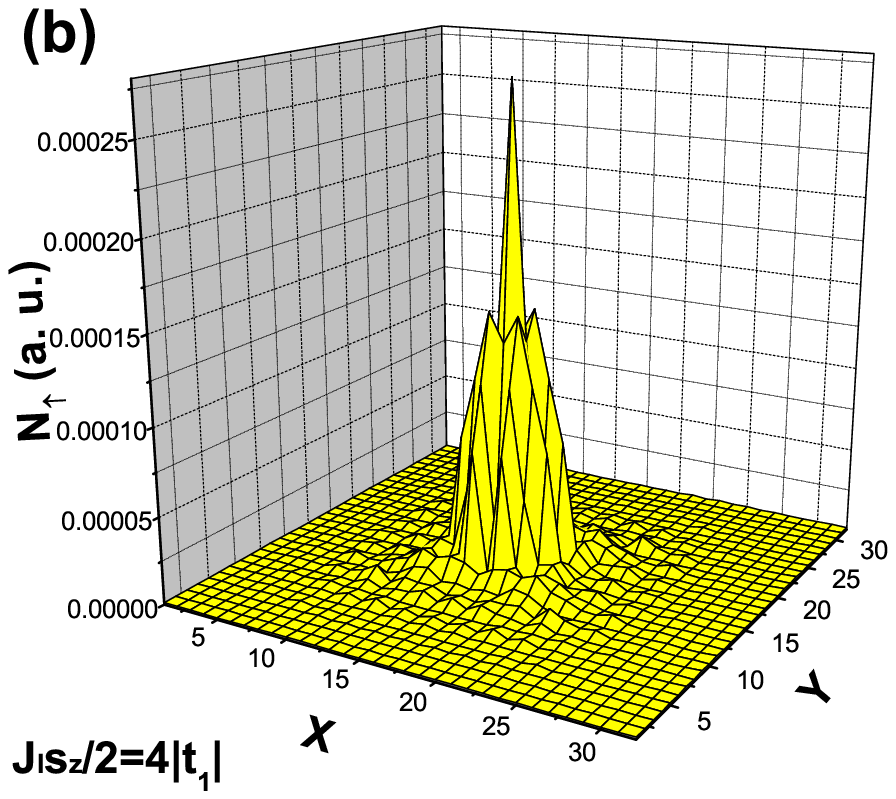}
\end{center}
\caption{(Color online) The spatial distribution of the spin-resolved
LDOS for $s_{\pm}$-wave pairing symmetry at $\omega=|\Omega_0|$ with (a) $J_Is_z/2=1.5$ before the transition, and (b) $J_Is_z/2=4$ after the transition. Only spin-down (spin-up) distribution is shown in (a) [(b)], while the others are nearly zero on the entire lattice.}\label{fig:spinLDOS}
\end{figure}

{\it The in-gap bound state is more robust for the $s_{\pm}$-wave pairing symmetry in the strong impurity scattering regime}.
In single-band $s$-wave superconductors, one of remarkable properties due to a localized magnetic impurity is that the first-order quantum phase transition takes place as the effective moment, $s^{eff}=J_Is_z/2$, is greater than certain critical value, $s^{eff}_c$. This transition represents the jump of the spin quantum number of the ground state from $0$, where the magnetic impurity is unscreened, to spin $1/2$, where the magnetic impurity is partially screened. We refer interested readers to the review paper by Balatsky {\it et al.} for detailed discussions.\cite{Balatsky2006} Here, in our case, one indication for such a transition is given by Eq.~(\ref{eq:mimpBS}), where $\Omega_0$ switches sign as $\frac{J_Is_z}{2}>\frac{1}{\pi\rho_0|\gamma_0|}$. This is similar to the case of $s$-wave state but with slightly different critical value. However, in the strong scattering limit, where $c_m\gg 1$, we observe that the bound-state energy for $s_{\pm}$-wave pairing never evolves back to the gap edge, but it does for the onsite $s$-wave pairing, as shown clearly in Figs.~\ref{fig:LocationBS}(a) and \ref{fig:LocationBS}(b), respectively, from our numerical study (flipping the sign of $c_m$ does not change the result). This feature may suggest the bound state solution in the sign-changed SC state is more robust than the one in the sign-unchanged case in the strong scattering regime. Note that the non-crossing to the zero energy is understood due to the local SC gap relaxation and many-body effect.\cite{Salkola97,Balatsky2006}

\begin{figure} [htbp]
\begin{center}
\includegraphics[width=0.42\textwidth]{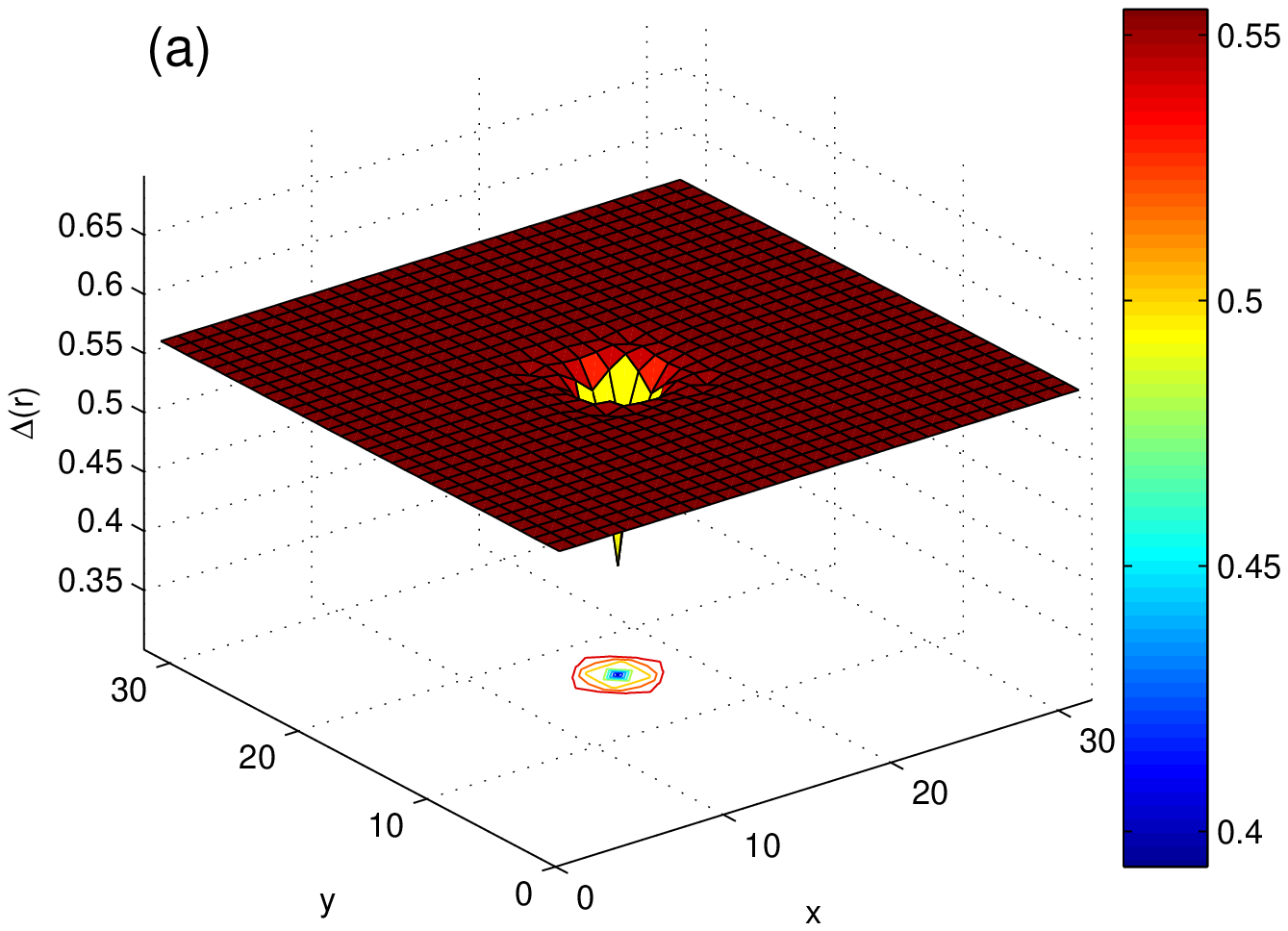}
\includegraphics[width=0.42\textwidth]{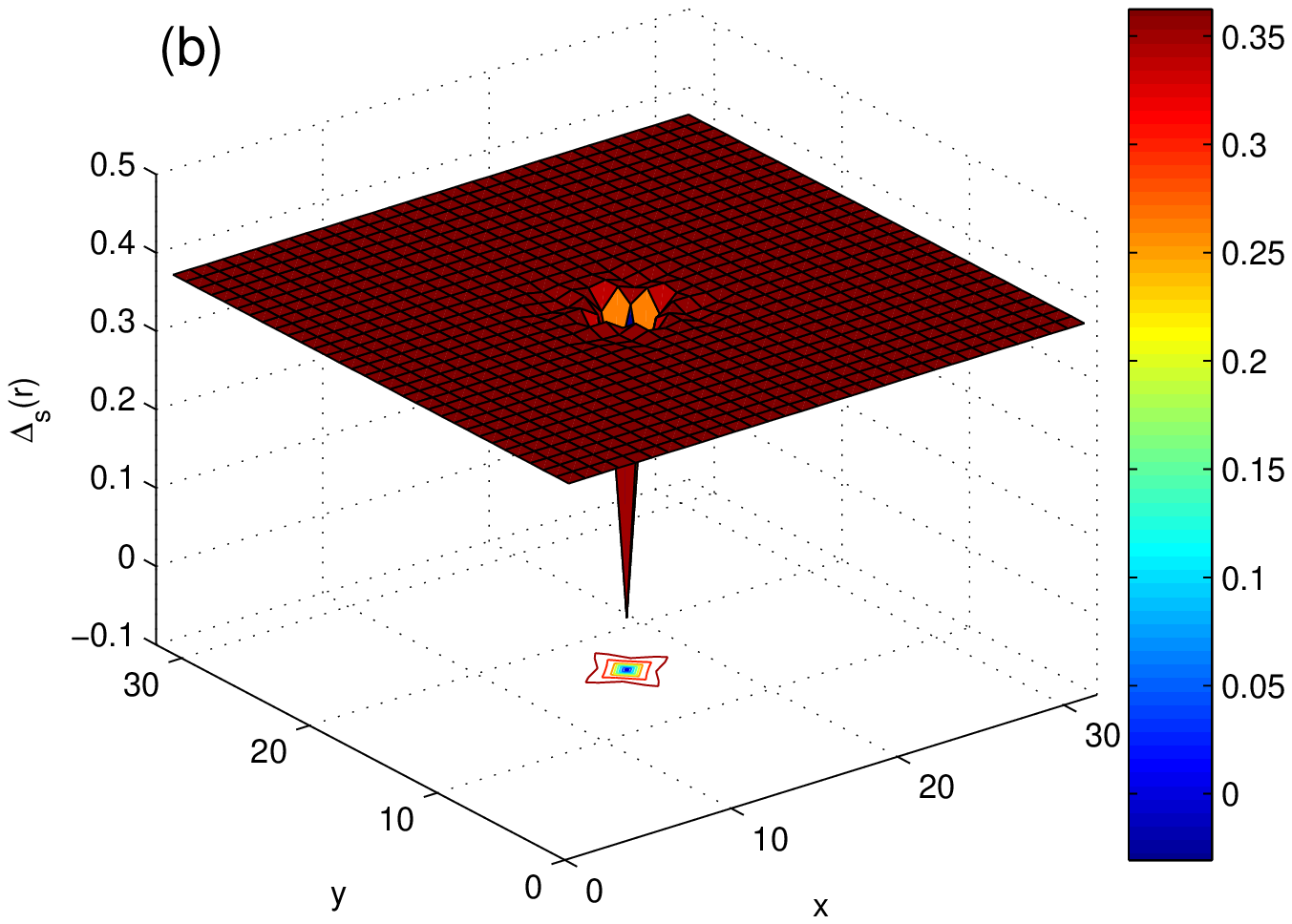}
\end{center}
\caption{(Color online) The spatial dependence of the SC gaps for (a) $s_{\pm}$-wave pairing symmetry and (b) the onsite $s$-wave pairing symmetry.
The parameters are same as those used in Fig.~\ref{fig:GraphLDOS}.} \label{fig:mimpSCgap}
\end{figure}

A further evidence to verify this transition (level crossing) can be provided by the spin-resolved LDOS, defined in Eq.~(\ref{eq:ldos}), around the impurity site. As shown in Figs.~\ref{fig:spinLDOS}(a) and \ref{fig:spinLDOS}(b), one can clearly see that the spin configurations are completely interchanged as $J_Is_z/2$ crosses the critical value$\sim 2.2$. Similar results are also found for the onsite $s$-wave pairing state (not shown).

\begin{figure} [htbp]
\begin{center}
\includegraphics[width=0.238\textwidth]{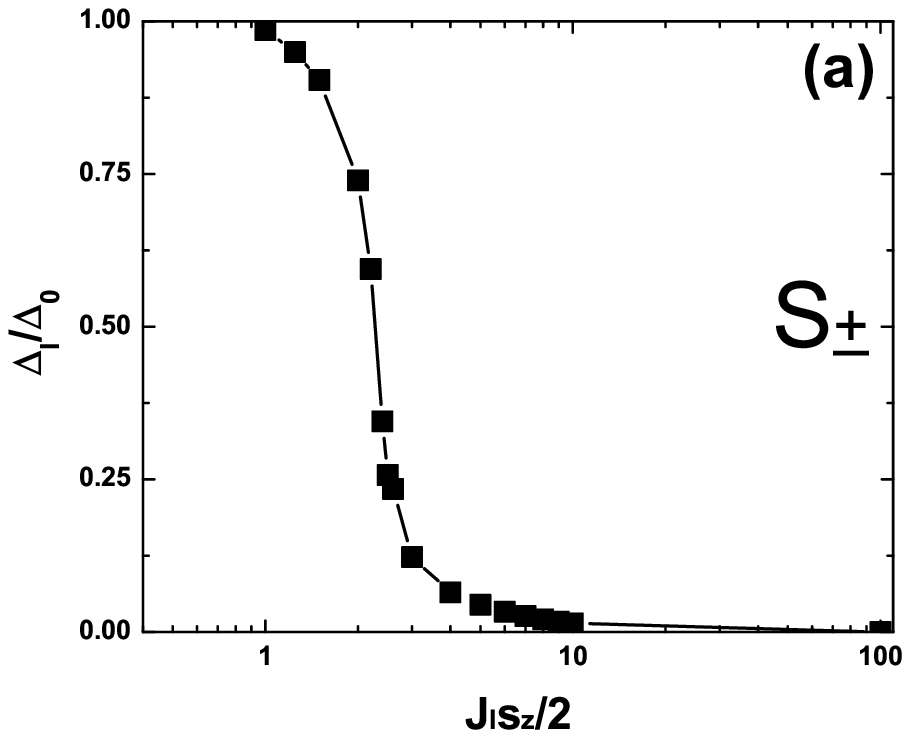}
\includegraphics[width=0.238\textwidth]{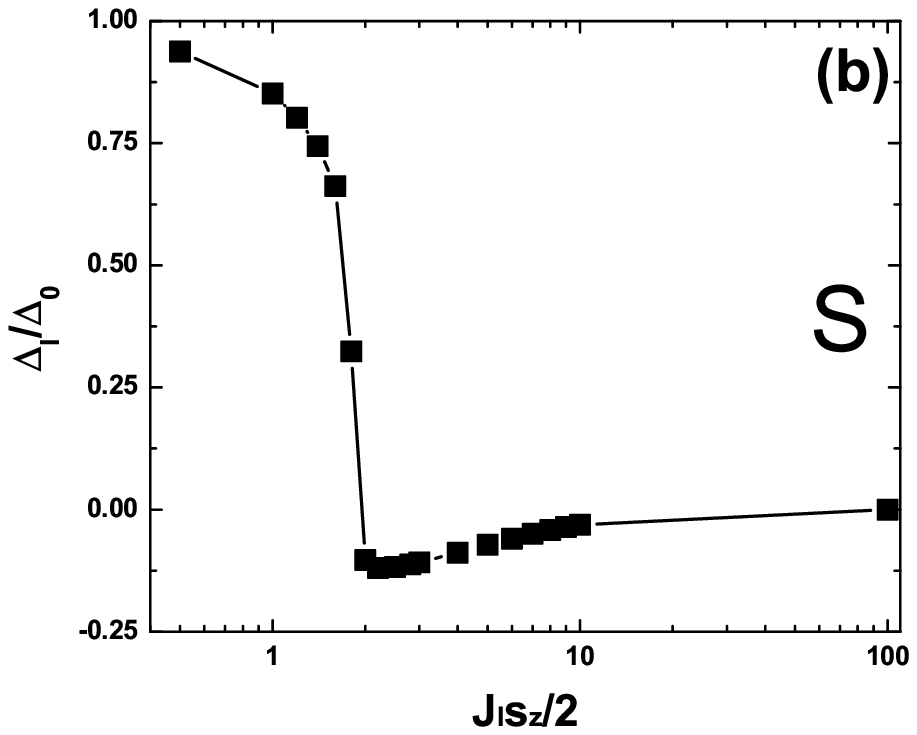}
\end{center}
\caption{The SC order parameter ($d_{xz}$ orbital only) at the impurity
site $\br_I$ as a function of the effective magnetic moment $J_Is_z/2$ for
(a) $s_{\pm}$-wave and (b) onsite $s$-wave. Note that $x$ axis is in logarithmic scale and the rescaling factor $\Delta_0$ of the SC gap is chosen to be 0.54 (0.38) for graph (a) [(b)].} \label{fig:DeltaI}
\end{figure}

{\it No $\pi$ phase shift of the SC gap function around the impurity for $s^{eff}>s_c^{eff}$ with the $s_{\pm}$-wave pairing symmetry}.
In Figs.~\ref{fig:mimpSCgap}(a) and \ref{fig:mimpSCgap}(b), we show the self-consistent SC pairing potentials for the $s_{\pm}$-wave and $s$-wave pairing symmetries, respectively. The basic features are not much different from the cases we have discussed in the non-magnetic impurity problem. However, as pointed out by Salkola {\it et al.},\cite{Salkola97} the phase of the SC order parameter changes by $\pi$ at the {\it magnetic}-impurity site with respect to the bulk phase when $s^{eff}$ is larger than the critical value. This is indeed a sharp feature we have seen for the sign-unchanged $s$-wave pairing state in Fig.~\ref{fig:DeltaI}(b), but not for the $s_{\pm}$-wave pairing state in Fig.~\ref{fig:DeltaI}(a).

Finally, we comment on the case when there exists small component of the inter-orbital impurity scattering, {\it i.e.}, $J_I\neq 0$ and $J_I^\prime\neq 0$. Similar to the case with a nonmagnetic impurity, this would lead to the splitting of the bound-state energy due to the breakdown of the orbital degeneracy. Consequently, we now have two critical values, $s^{eff}_{c1}$ and $s^{eff}_{c2}$, when increasing the effective magnetic moment, $J_Is_z/2$ with fixed ratio $J_I^\prime/J_I$. These transitions correspond to the change in the spin quantum number in the ground state, that is, from 0 to 1/2 and 1/2 to 1, respectively. When $s^{eff}_{c1}<s^{eff}<s^{eff}_{c2}$, the spin and orbital degrees of freedom are strongly correlated and more complicated spin configurations are expected. This may deserve further study in the future.

\section{Remarks on nodal pairing symmetries and many impurities}
Although we have mainly concentrated on the effects of a {\it single} nonmagnetic (magnetic) impurity in {\it fully gapped}, sign-changed, $s$-wave superconductors, we would like to make two remarks which are closely related to, or suggested by the current study.

(1) Considering the case of sign changed, but with SC nodes on electron Fermi pockets, as called for attention in recent theoretical works,\cite{Graser09,Maier2009a,chubukov2009} is of course beyond the scope of this paper and may deserve a future study. However, following the same $T$-matrix calculations sketched above, we can consider certain nodal pairing states in a straightforward manner. For instance, the gap function could be of the form, $\Delta_{\alpha}(\bk)=\Delta_0(\cos k_x+\cos k_y)/2$, {\it i.e.}, the $A_{1g}$ nodal pairing symmetry with nodes on the electron pockets. When evaluating $g^0(\tilde{\omega})$, one can realize that the $\tau_1$ component in Eq.~(\ref{eq:tinverse}) is not absent (or negligible) anymore (with contribution mainly from the hole pockets), and it makes the system much closer to a sign-unchanged $s$-wave pairing state. Thus, in the single nonmagnetic-impurity problem, the should-be-present peaks in the LDOS disappear or are nearly indistinguishable with the gap edge (continuum).

(2) There are already several papers\cite{Bang09,Senga08,Senga09} discussing about the issue of many impurities and its relation to the SC $T_c$ with sign-changed $s$-wave pairing. In particular, Senga and Kontani\cite{Senga09} present a detailed study on the (nonmagnetic) impurity-induced DOS and the suppressed $T_c$ with various {\it inter-band} ({\it intra-band}) scattering strengths $I^\prime$ ($I$) within $T$-matrix approach.
According to their results, at the fixed ratio $I^\prime/I=1$, the induced-impurity band would move toward zero energy without going back to the gap edge as increasing the scattering strength, associated as well with a large suppression of $T_c$. This tendency corresponds to our observation that the impurity-induced bound-state energy never evolves back to the gap edge as increasing the scattering strength, and should be in sharp contrast to the sign-unchanged $s$-wave pairing state, in which Anderson's theorem\cite{Anderson59} is expected to be satisfied.

\section{Conclusions}
In conclusion, we have studied the impurity-induced in-gap bound states in iron-based superconductors with (sign-changed) $s_{\pm}$-wave pairing symmetry by using both the self-consistent BdG formulation and non-self-consistent $T$-matrix approach. In comparison to the sign-unchanged $s$-wave pairing state, we have found several signatures, which are mainly associated with the sign change in the SC order parameter. In particular, for a nonmagnetic impurity, the two in-gap bound-state peaks appear in the LDOS at or near the impurity site and their formation is due to the sign-reversal effect in the order parameter during Andreev reflection processes.
For a magnetic impurity, there also exist bound-state solutions, but only for one of the electron-spin polarizations around the impurity at the resonance energy due to the breakdown of the local time-reversal symmetry. Above a critical effective magnetic moment, the ground state of the system undergoes a quantum phase transition, from a spin-unpolarized state to a spin-polarized one. Although in the presence of a magnetic impurity, both sign-changed and sign-unchanged $s$-wave pairing states behave qualitatively the same, we emphasize that the former pairing state can sustain more robust bound-state solutions without a $\pi$ phase shift of the SC gap near the impurity in the strong scattering regime.

{\it Note added-}
As we nearly complete our paper, we find two interesting papers by Zhang,\cite{zhang2009b} and Tao Zhou {\it et al.},\cite{zhou2009d} discussing similar issues on the impurity effect. Their insightful results are basically consistent with ours.

\begin{acknowledgements}
J.P.H. thanks B. A. Bernevig for useful
discussions.  J.P.H., Y.Y.Z., W.F.T., and C.F. were supported by the
NSF under Grant No. PHY-0603759.
\end{acknowledgements}

\appendix
\section{Derivation of $g^0(\tilde{\omega})$ within T-matrix approximation}
In order to calculate the bare Green's function $G^0(\bk,\tilde{\omega})$ and its corresponding $g^0(\tilde{\omega})$, it is more convenient to turn our orbital basis into band representation, where we can easily obtain the Green's function for each band, by the following unitary transformation
\begin{eqnarray}
U^\dag(\mathbf{k})\left(
\begin{array}{cc}
  \epsilon_x(\mathbf{k}) & \epsilon_{xy}(\mathbf{k}) \\
  \epsilon_{xy}(\mathbf{k}) & \epsilon_y(\mathbf{k}) \\
\end{array}
\right)U(\mathbf{k})=\left(
\begin{array}{cc}
  \epsilon_e(\mathbf{k}) & 0 \\
  0 & \epsilon_h(\mathbf{k}) \\
\end{array}
\right),
\end{eqnarray}
where
\ba U(\bk)&=&\left(
\begin{array}{cc}
  \cos(\theta_{\mathbf{k}}/2) & -\sin(\theta_{\mathbf{k}}/2) \\
  \sin(\theta_{\mathbf{k}}/2) & \cos(\theta_{\mathbf{k}}/2) \\
\end{array}
\right), \\
\cos{\theta_k}&=&\frac{\epsilon_-(\bk)}
{\sqrt{\epsilon^2_-(\bk)+\epsilon^2_{xy}(\bk)}},
\sin{\theta_k}=\frac{\epsilon_{xy}(\bk)}
{\sqrt{\epsilon^2_-(\bk)+\epsilon^2_{xy}(\bk)}}, \nonumber
\ea
and $\epsilon_{x(y)}(\mathbf{k})=\epsilon_+(\mathbf{k})\pm\epsilon_-(\mathbf{k})$. In the normal state, $\epsilon_{e(h)}(\mathbf{k}_f)=\mu$ associates with two electron (hole) pockets. $H_0^{MF}$ now transforms as $\tilde{H}_0^{MF}=\sum_\mathbf{k}
\tilde{\Psi}^{\dagger}(\mathbf{k})\tilde{h}(\mathbf{k})\tilde{\Psi}(\mathbf{k})$,
where $\tilde{h}(\mathbf{k})=\{[(\epsilon_e(\mathbf{k})+\epsilon_h(\mathbf{k}))/2 -\mu]\sigma_0+[(\epsilon_e(\mathbf{k})-\epsilon_h(\mathbf{k}))/2]\sigma_3
\}\otimes\tau_3+\Delta(\mathbf{k})\sigma_0\otimes\tau_1$.
Consequently, the bare Green's function $\tilde{G}^{0}(\bk,\tilde{\omega})$ for band electrons is given by
\be
[(\omega+i0^+)I_4-\tilde{h}(\bk)]^{-1}\equiv
\left(\begin{array}{cc}
  \tilde{G}^{0}_{e}(\bk,\tilde{\omega}) & 0 \\
0 & \tilde{G}^{0}_{h}(\bk,\tilde{\omega})
\end{array}
\right),
\ee
where
\be
\tilde{G}^{0}_{e(h)}(\bk,\tilde{\omega}) = \frac{\tilde{\omega}\tau_0+\Delta(\bk)\tau_1+(\epsilon_{e(h)}(\bk)-\mu)\tau_3}
{\tilde{\omega}^2-(\epsilon_{e(h)}(\bk)-\mu)^2-\Delta^2(\bk)}.
\ee
Now, we can transform above Green's function back to its orbital representation defined in Eq.~(\ref{eq:G0}) with matrix elements,
\ba
G^0_{11(22)}(\bk,\tilde{\omega})&=&\tilde{G}^0_{e(h)}(\bk,\tilde{\omega})
\cos^2(\frac{\theta_{\bk}}{2}) \nonumber \\
&+&\tilde{G}^0_{h(e)}(\bk,\tilde{\omega})
\sin^2(\frac{\theta_{\bk}}{2}),  \\
G^0_{12(21)}(\bk,\tilde{\omega})&=&
[\tilde{G}^0_{e}(\bk,\tilde{\omega})
-\tilde{G}^0_{h}(\bk,\tilde{\omega})]\sin(\frac{\theta_{\bk}}{2})
\cos(\frac{\theta_{\bk}}{2}).\nonumber
\ea
%\end{widetext}

\begin{figure}[bth]
\begin{center}
\includegraphics[scale=0.4]{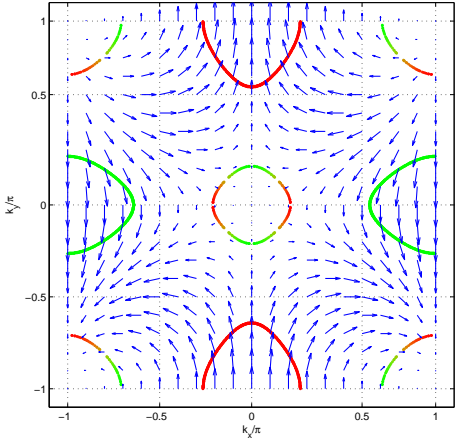}
\caption{The arrows show the direction and magnitude of the vector $(\epsilon_{xy}(\bk),\epsilon_-(\bk))$ in the (unfolded) FBZ. It is then easy to extract $\theta_{\bk}$ information from this arrow map. The orbital-resolved Fermi pockets are also put on top of the map for convenience. Red and green colors represent $d_{xz}$ and $d_{yz}$ orbitals, respectively. ($\mu=1.65$)}
\label{fig:thetamap}
\end{center}
\end{figure}

Upon integrating over momentum in the FBZ, we obtain the following approximate expression for $g^0(\tilde{\omega})$,
%\begin{widetext}
\ba
g^0(\tilde{\omega})&=&\int\frac{d^2k}{(2\pi)^2}
\left(\begin{array}{cc}
  G^{0}_{11}(\bk,\tilde{\omega}) & G^{0}_{12}(\bk,\tilde{\omega}) \\
G^{0}_{21}(\bk,\tilde{\omega}) & G^{0}_{22}(\bk,\tilde{\omega})
\end{array}
\right), \nonumber \\
&=& \frac{1}{2}\int\frac{d^2k}{(2\pi)^2}
\sigma_0\otimes\left(\tilde{G}^0_{e}(\bk,\tilde{\omega})
+\tilde{G}^0_{h}(\bk,\tilde{\omega})\right)
,\nonumber \\
&\approx&
-\pi\rho_0\sigma_0\otimes\left(\frac{\alpha(\tilde{\omega})}{\sqrt{\Delta_0^2-\tilde{\omega}^2}}
\tilde{\omega}\tau_0-\gamma(\tilde{\omega})\tau_3\right),
\, (\text{$s_{x^2y^2}$}) \nonumber  \\
&\text{or}&
-\pi\rho_0\sigma_0\otimes\left(\frac{\alpha(\tilde{\omega})
(\tilde{\omega}\tau_0+\Delta_0\tau_1)}{\sqrt{\Delta_0^2-\tilde{\omega}^2}}
-\gamma(\tilde{\omega})\tau_3\right),
\, (\text{$s$})\nonumber \\
\ea
%\end{widetext}
where $\alpha(\tilde{\omega})= \frac{1}{\pi}[\tan^{-1}\left(\frac{E_c}
{\sqrt{\Delta^2-\tilde{\omega}^2}}\right)+
\tan^{-1}\left(\frac{E_g}
{\sqrt{\Delta^2-\tilde{\omega}^2}}\right)]\sim 1$, $\gamma(\tilde{\omega})= \frac{1}{2\pi}\ln\left(\frac{E_c^2+\Delta^2-\tilde{\omega}^2}
{E_g^2+\Delta^2-\tilde{\omega}^2}\right)\sim\gamma_0$, with $E_g$ representing the average energy difference between the pocket center and the Fermi level and $E_c$ representing the energy cutoff with respect to the Fermi level.\cite{Tsai09} $\rho_0$ is the density of states at the Fermi level.

Note that to get the second equality, we have taken into account the features of $\theta_{\bk}$ in the FBZ (see Fig.~\ref{fig:thetamap}); to get the third/fourth equality, we have made several approximations: (i) for each pocket, the energy dispersion is quadratic with respect to the pocket center, (ii) the density of states around the Fermi level (and within the cutoff energy $E_c$) for each pocket is a constant $\rho_0$, and (iii) for $s_{\pm}$-wave pairing, $\Delta(\bk)\sim\Delta_{coh}$ for hole pockets while $-\Delta_{coh}$ for electron pockets. This is in contrast to the on-site $s$-wave case where we take $\Delta(\bk)=\Delta_{coh}$ for all Fermi pockets.\cite{Tsai09}


\begin{thebibliography}{99}
\bibitem{comment_tau1}
{\sc T}he presence of a small piece of $\tau_1$ in $T^{-1}(\tilde{\omega})$
  will not change our results qualitatively, but will shift the critical value
  of $s^{eff}$ to reach zero energy in the $T$-matrix approximation.
\end{thebibliography}
\end{document}